\documentclass[letterpaper,twocolumn,10pt]{article}

\usepackage[most]{tcolorbox}
\usepackage{usenix}

\usepackage[available]{usenixbadges}

\usepackage{tikz}
\usepackage{svg}
\usepackage{amsmath}
\usepackage{graphicx}

\usepackage{subcaption}
\usepackage{enumitem}
\usepackage{tabularx}

\usepackage{booktabs}
\usepackage{xspace}
\usepackage[export]{adjustbox}

\usepackage{authblk}

\usepackage{hyperref}
\usepackage{cleveref}

\mathchardef\UrlBreakPenalty=0
\mathchardef\UrlBigBreakPenalty=0

\usepackage{customenvs}

\usepackage{msc}
\usepackage{wasysym}
\usepackage{amssymb}
\usepackage{pifont}

\usepackage{amsthm}
\newtheorem{definition}{Definition}

\usepackage[colorinlistoftodos,prependcaption,disable]{todonotes} %
\newcommand{\aju}[1]{\todo[linecolor=blue,backgroundcolor=blue!25,bordercolor=blue,inline]{\textbf{Aljosha:} #1}\noindent}

\newcommand{\dsc}[1]{\todo[linecolor=orange,backgroundcolor=orange!25,bordercolor=orange,inline]{\textbf{David:} #1}\noindent}

\newcommand{\iWhatsApp}{\includegraphics[height=1em]{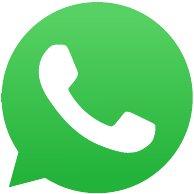}}%
\newcommand{\iSignal}{\includegraphics[height=1em]{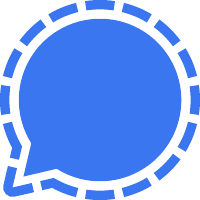}}%
\newcommand{\iThreema}{\includegraphics[height=1em]{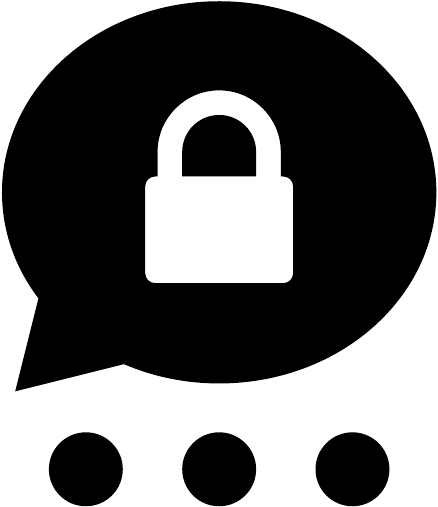}}%
\newcommand{\iMessage}{\includegraphics[height=1em]{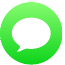}}%

\newcommand{\deviceAck}{\includegraphics[height=0.8em]{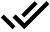}}%
\newcommand{\readAck}{\includegraphics[height=1em]{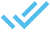}}%

\newcommand{\eThumbup}{\includegraphics[height=1em]{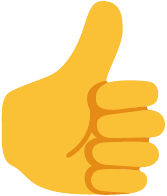}}%

\usepackage{tikz}
\usepackage{amsmath}
\usepackage{xcolor}
\newcommand{\roundframe}[1]{{\setlength\fboxrule{0pt}\fbox{\tcbox[colframe=black,colback=white,shrink tight,boxrule=0.5pt,extrude by=2.5pt]{\small #1}}}}

\tcbset{
  takeawaystyle/.style={
    colback=gray!5!white, %
    colframe=gray!75!black, %
    fonttitle=\bfseries\itshape, %
    title={#1}, %
    attach boxed title to top left={xshift=2mm, yshift=-3mm}, %
    boxed title style={
      colframe=white, %
      boxrule=0mm, %
      boxsep=1mm, %
    },
    enhanced, %
    boxrule=0.5mm, %
    top=3mm, %
  },
}

\usepackage{listings}
\lstdefinelanguage{json}{
  sensitive=true,
  morestring=[b]",
  alsoletter={:},
  morekeywords={true,false,null},
  literate=*%
    {0}{{{\color{blue}0}}}{1}
    {1}{{{\color{blue}1}}}{1}
    {2}{{{\color{blue}2}}}{1}
    {3}{{{\color{blue}3}}}{1}
    {4}{{{\color{blue}4}}}{1}
    {5}{{{\color{blue}5}}}{1}
    {6}{{{\color{blue}6}}}{1}
    {7}{{{\color{blue}7}}}{1}
    {8}{{{\color{blue}8}}}{1}
    {9}{{{\color{blue}9}}}{1}
}

\newcommand{\greenCheck}{\textcolor{red}{\faCheck}\xspace}     
\newcommand{\redRemove}{\textcolor{black}{\faTimes}\xspace} 
\newcommand{\neutralMinus}{\textcolor{gray}{\faMinus}\xspace}

\usepackage[nohyperlinks, nolist]{acronym}

\begin{acronym}
  \acro{E2EE}[E2EE]{End-to-end encrypted}
  \acro{TC}[TC]{transcript consistency}
  \acro{OTR}[OTR]{Off-the-Record Messaging}
  \acro{mpOTR}[mpOTR]{multi-party Off-the-Record Messaging}
  \acro{DS}[DS]{delivery service}
  \acro{PoC}[PoC]{Proof of Concept}

\end{acronym}

\usepackage{xcolor}
\tcbuselibrary{skins,breakable}

\definecolor{whatsappdark}{HTML}{075E54}
\definecolor{whatsappgreen}{HTML}{DCF8C6}
\definecolor{whatsappgray}{gray}{0.45}
\definecolor{whatsappdelivered}{HTML}{53bdeb}

\newtcolorbox{whatsappbubble}[1][]{
  enhanced,
  colback=whatsappgreen,
  colframe=whatsappgreen,
  boxrule=0pt,
  arc=3mm,
  left=2mm,
  right=3mm,
  top=2mm,
  bottom=6mm, %
  width=\linewidth,
  flush right,
  after skip=2mm,
  clip upper=false,
  clip lower=false,
  overlay={
    \path[fill=whatsappgreen]
      ([xshift=-6mm,yshift=0mm]frame.north east) --
      ([xshift=4mm,yshift=0mm]frame.north east) --
      ([xshift=0mm,yshift=-6mm]frame.north east) -- cycle;

    \node[
      anchor=south east,
      font=\bfseries\small,
      text=whatsappdelivered
    ] at ([xshift=-1.5mm,yshift=1.5mm]frame.south east) {Takeaway {\large\readAck}};
  },
}

\newenvironment{whatsapptakeaway}[1][]{%
  \begin{whatsappbubble}[#1]%
}{%
  \end{whatsappbubble}%
}

\begin{document}

\hyphenation{macOS}
\hyphenation{iOS}
\hyphenation{WhatsApp}

\date{}

\title{\Large \bf 
Send and Pretend:
Exploiting Transcript Consistency Issues in End-to-End Encrypted Group Chats
}

\makeatletter
\renewcommand\AB@affilsepx{, \protect\Affilfont}
\makeatother

\author[1]{Gabriel K. Gegenhuber}
\author[2]{Moritz Grefner}
\author[3]{Maximilian Günther}
\author[2]{Matthäus Wininger}
\author[2,4,5]{\authorcr David Schmidt}
\author[2]{Aljosha Judmayer}

\affil[1]{Interdisciplinary Transformation University (IT:U)}
\affil[2]{University of Vienna, Faculty of Computer Science}
\affil[3]{SBA~Research}
\affil[4]{UniVie Doctoral School Computer Science}
\affil[5]{CDL AsTra}

\maketitle

\begin{abstract}
End-to-end encrypted (E2EE) messaging apps are widely praised for their security and thus also used for sensitive coordination in group chats (e.g., by political decision makers). 
After Threema and WhatsApp, also Signal and iMessage have recently introduced polls to aid agreement processes in groups. This implicitly sets the expectation that all participants see the same outcome and thus have the same view of the conversation. 
This property is commonly referred to as \emph{transcript consistency} (TC).
In this work, we demonstrate that today's major E2EE messengers do not guarantee \emph{any} form of TC for group chats, allowing a malicious group member to selectively omit, reorder, or present altered content to different recipients without triggering warnings in their user interface.
We systematically investigate the extent of the problem under a \emph{malicious-participant} threat model that targets the integrity of the shared transcript, or inconsistent delivery across a user's linked devices.
We identify multiple equivocation vectors that range from protocol fallback paths to deliberate use of pairwise delivery channels within groups.
We demonstrate concrete exploitation scenarios such as social engineering, evading moderation, and, in particular, \emph{rigging polls}.
Beyond these cross-service design issues, we also uncover implementation-specific behaviors with privacy implications (e.g., device OS fingerprinting).
Finally, we contextualize our findings within prior transcript-consistency research and outline practical low-overhead mitigations and UI signaling strategies that can be integrated into state-of-the-art E2EE group protocols.

\end{abstract}

\section{Introduction}

\begin{figure*}[htb]
    \centering
    \begin{subfigure}[b]{0.33\textwidth}
        \centering
        \includegraphics[fbox=0.5pt 0pt,trim={0 35cm 0 0},clip,width=\linewidth]{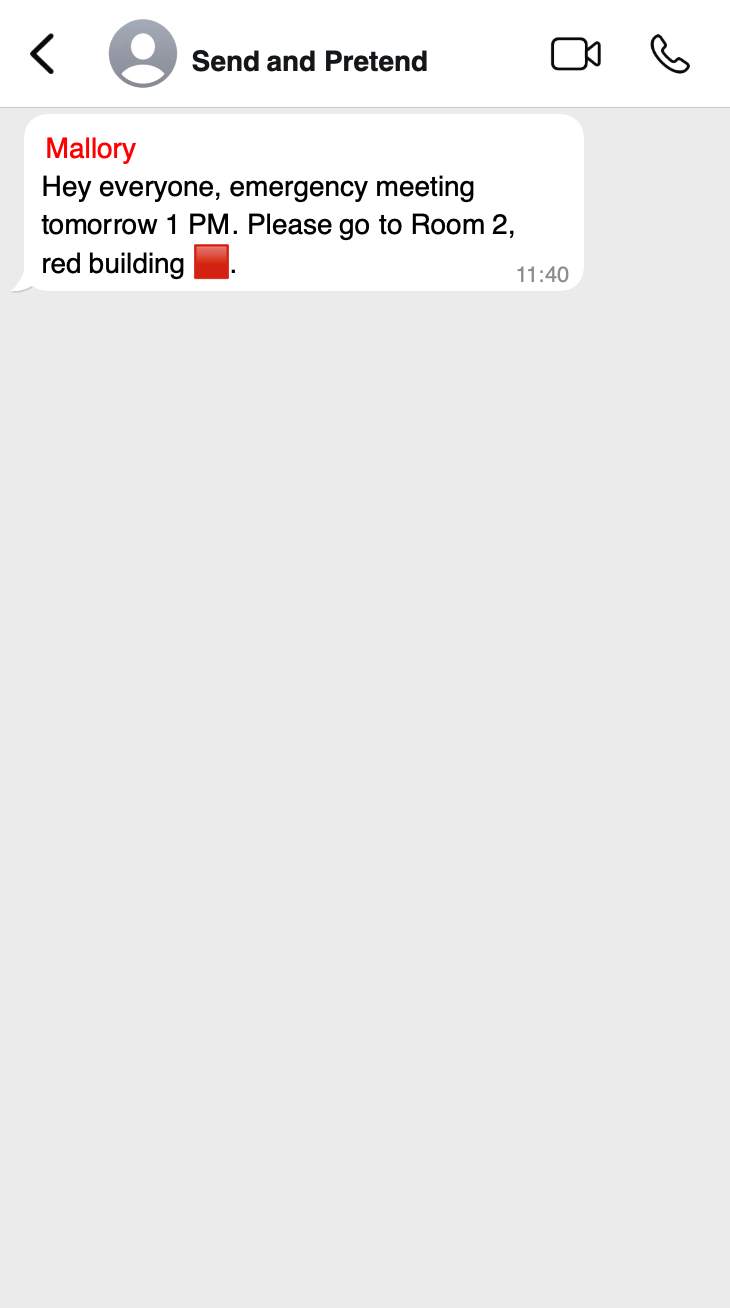}
        \caption{Alice (original message)}
        \label{fig:chat-alice}
    \end{subfigure}
    \hfill
    \begin{subfigure}[b]{0.33\textwidth}
        \centering
        \includegraphics[fbox=0.5pt 0pt,trim={0 35cm 0 0},clip,width=\linewidth]{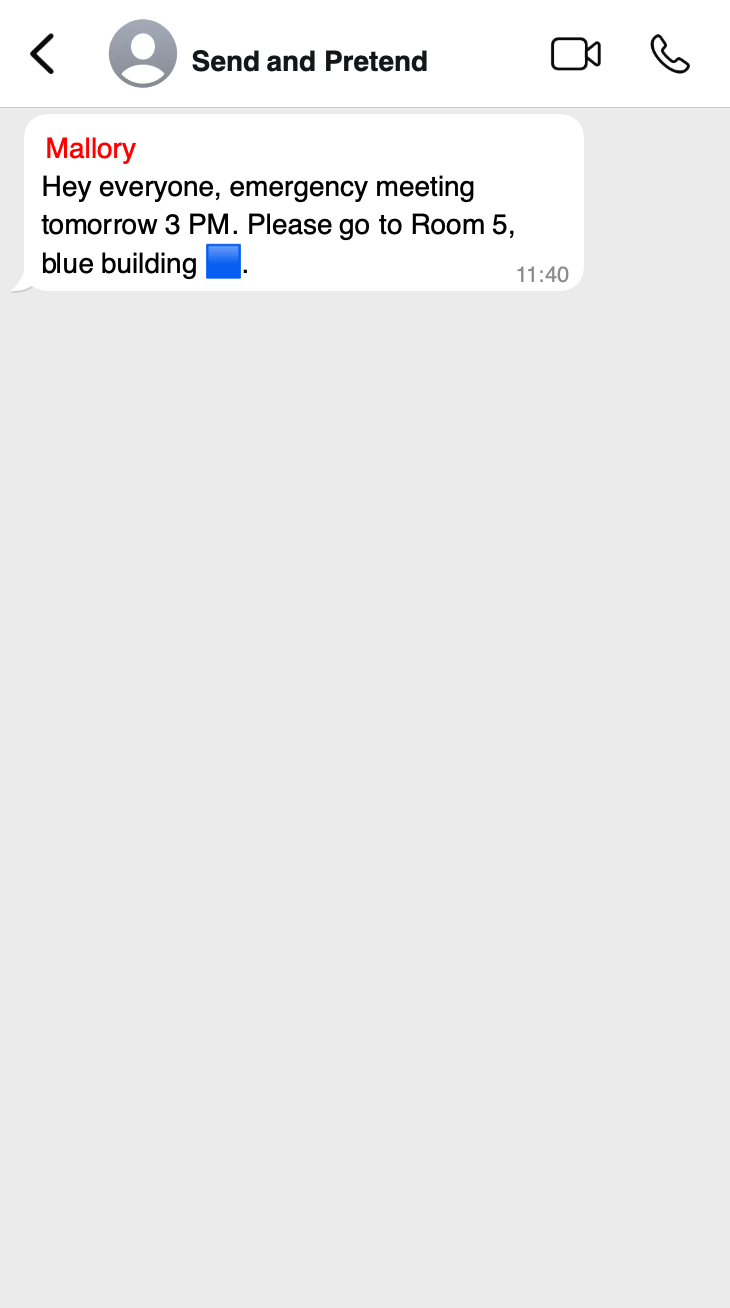}
        \caption{Bob (victim)}
        \label{fig:chat-bob}
    \end{subfigure}
    \hfill
    \begin{subfigure}[b]{0.33\textwidth}
        \centering
        \includegraphics[fbox=0.5pt 0pt,trim={0 35cm 0 0},clip,width=\linewidth]{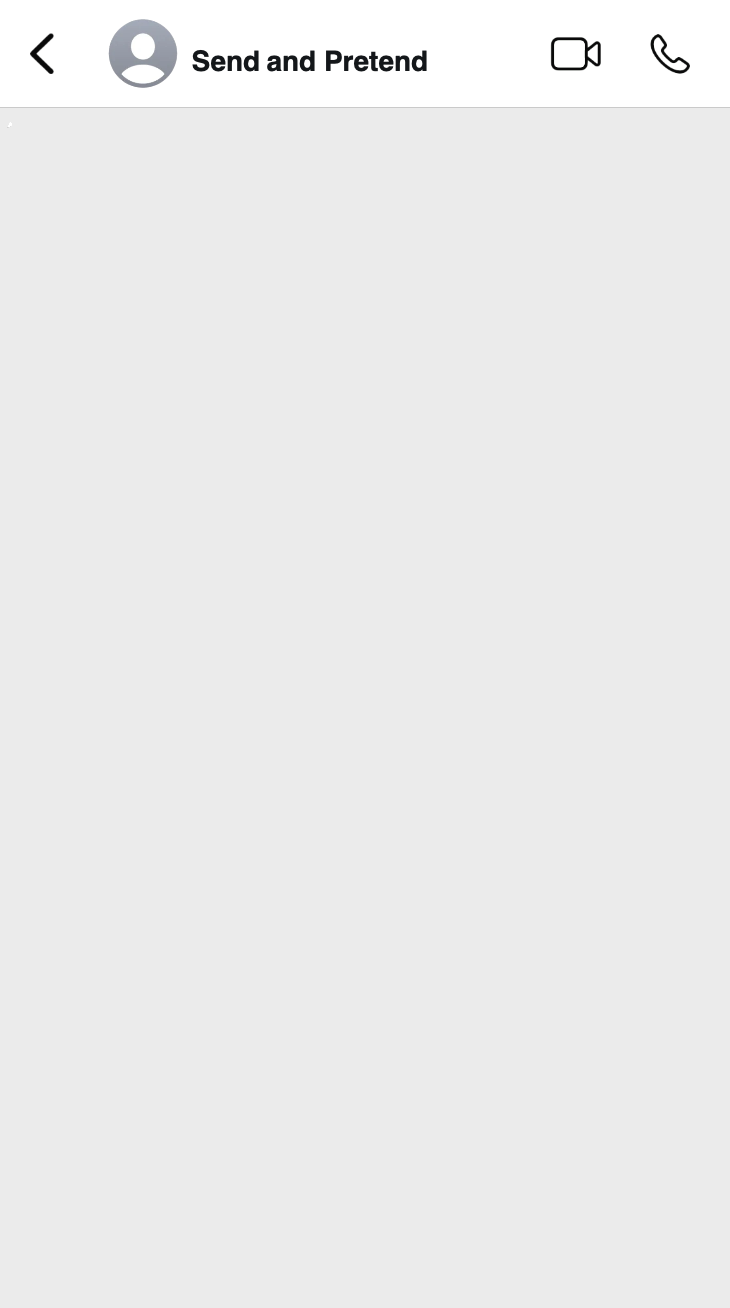}
        \caption{Carol (victim)}
        \label{fig:chat-carol}
    \end{subfigure}

    \caption{Side-by-side views of a group chat from each participant's perspective. 
    Each panel represents the local transcript on a participant’s device. While Alice sees the original message, Bob receives a message with an exchanged meeting time, room, and building, and Carol was skipped entirely.
    At present, this attack is possible across all prevalent end-to-end encrypted (E2EE) messengers without being detected by the server or triggering warnings on the receiving client.
    }
    \label{fig:whatsapp-side-by-side}
\end{figure*}

\ac{E2EE} instant messaging applications such as Signal, WhatsApp, iMessage, and Threema are used daily by a substantial fraction of the global population~\cite{gegenhuber_2025_heythere,mau_imessage,mau_signal,mau_threema}. 
A precise assessment of the security guarantees provided by these apps is therefore critical for private individuals, as well as government officials, particularly in the United States~\cite{goldberg_trump_2025,lee_despite_2025} and the European Union~\cite{holland_eu_2025}.
In these regions, encrypted messaging apps are reportedly used for sensitive governmental communication and decision-making, as illustrated by a recent incident in which a journalist was erroneously added to a group chat of government officials, including the US Secretary of War, Pete Hegseth~\cite{goldberg_trump_2025}.

A key security property in this context is that all members of a group conversation eventually converge to the same view of the message transcript.
This property is commonly referred to as \emph{\ac{TC}}\footnote{Comparable security properties were discussed in academia under different names, e.g., \emph{consensus}~\cite{goldberg_multi-party_2009}, \emph{speaker consistency} and \emph{global transcript}~\cite{unger_sok_2015}, \emph{causality preservation}~\cite{unger_sok_2015,chen_integrating_2024}, \emph{transcript agreement}~\cite{cohn-gordon_ends--ends_2018}, \emph{transcript consistency}~\cite{marlinspike_private_2014,kleppmann_secure_2018}.}.
In the evolution of \ac{E2EE} instant messaging, \ac{TC} has been proposed as a desirable property for group chats early on~\cite{goldberg_multi-party_2009} and was also discussed in the design of the TextSecure protocol~\cite{marlinspike_private_2014}, which later evolved into Signal and shaped the design of later messengers such as WhatsApp.

Prior work on secure group messaging primarily considered adversaries that control the server and/or the network, and addresses threats such as message dropping, reordering, or traffic analysis by outsiders~\cite{rosler_more_2018,chase_signal_2020,signal_messenger_new_2020}.
Early proposals already discussed retrospective detection of transcript inconsistencies, for example, in \ac{OTR} and its first group extensions~\cite{goldberg_multi-party_2009,marlinspike_private_2014,unger_sok_2015}.

Despite these early discussions of \ac{TC} as a requirement for \ac{E2EE} messaging systems, the extent to which modern messengers actually provide \ac{TC} remains underexplored.
This gap becomes more pronounced as messengers increasingly integrate interactive group features such as polls~\cite{signal_polls,threema_polls,whatsapp_2023_polls,whatsapp_2022_polls,imessage_polls}. 
Unlike ordinary messages, polls implicitly assume not only a shared transcript but also a common interpretation of the shared state, thereby amplifying the impact of even subtle inconsistencies.
These features, therefore, require a consistent view of the outcome and motivate us to study: \emph{RQ: What is the current state of \ac{TC} in \ac{E2EE} messaging?}

In contrast to prior work, we explicitly consider a malicious client, already participating in the group, that aims to manipulate the view on the conversation of selected participants.
This assumption is especially relevant in large groups, where honest behavior from all participants cannot be taken for granted\footnote{The Sender Key protocol, as used by WhatsApp and Signal, supports groups of around 1,000 participants~\cite{signal_support_group_chats,signalandroid_remoteconfig_2026}.}.
In such settings, an attacker could, for example, distribute malicious links to a subset of participants while sending benign links to others, thereby evading moderation.
More aggressively, attackers could steer opinions or inflict social harm by selectively excluding participants or by providing them with incorrect information, for example, about scheduled meetings, as demonstrated in Figure~\ref{fig:whatsapp-side-by-side}.
The widespread use of instant messengers and group chats by government officials and world leaders further amplifies the severity of this threat.

In this paper, we show that, contrary to common assumptions, current prevalent instant messengers provide no form of \ac{TC} for group chats.
We demonstrate that WhatsApp~\cite{whatsapp}, Signal~\cite{signal}, iMessage~\cite{iMessage}, and Threema~\cite{threema} allow malicious group members to selectively omit, reorder, or tailor messages for individual participants without detection by the messaging server or the recipient.
These weaknesses extend to higher-level features, enabling attackers to manipulate polls by presenting different outcomes to different participants, biasing results, or overwriting polls entirely.
Such attacks directly target the integrity of the shared group transcript.

To the best of our knowledge, we are the first to expose the full extent of this problem, in particular for targeted messages and interactive features such as polls, and to show that it represents a fundamental design flaw affecting all prevalent \ac{E2EE} instant messenger designs.
Given this negative result, we analyze the underlying causes, survey prior \ac{TC} research, map the attack surface, and propose practical changes that can improve \ac{TC} in prevalent secure messaging applications.

In detail, we make the following contributions: 
\begin{itemize} %
\item We show that WhatsApp, Signal, iMessage, and Threema, in their current versions, do not provide \emph{any} form of \ac{TC} in group chats, allowing malicious participants to omit, reorder, or modify messages for individual group members without detection. 
\item We demonstrate that the poll feature of all analyzed messengers is also affected, allowing poll creators and voters to manipulate the outcome.
This ranges from honest group participants seeing different outcomes, to manipulating the outcome towards a desired decision or  overwriting the poll result with a desired outcome. 
\item We uncover implementation-specific issues with implications for \ac{TC} and user privacy.
\item We discuss mitigations that require minimal changes to the Sender Key protocol used by Signal and WhatsApp, enabling the detection of potential attacks and the ability to warn users.
\end{itemize}

\section{Background}
\label{sec:background}

For group messaging, \ac{E2EE} messengers (in particular, those based on the Signal protocol) propose two different paradigms for sending and distributing messages to group participants.

\subsection{Pairwise Messaging (Client Fan-Out)}
\begin{figure*}[ht]
    \centering
        \centering
        \begin{subfigure}[b]{0.35\linewidth}
            \centering
            \includegraphics[width=\linewidth,page=1]{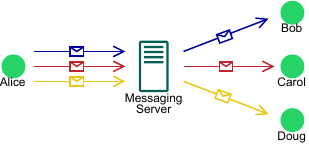}
            \caption{Client Fan-Out}
            \label{fig:client-fanout}
        \end{subfigure}
        \hspace{15ex}
        \begin{subfigure}[b]{0.35\linewidth}
            \centering
            \includegraphics[width=\linewidth,page=2]{figures/Client_Server_Fanout.drawio.pdf}
            \caption{Server Fan-Out}
            \label{fig:server-fanout}
        \end{subfigure}
    \caption{Comparison of client-side and server-side fan-out message delivery models.}
    \label{fig:fanout-models}
\end{figure*}

In the client fan-out model, group messages are handled similarly to one-to-one (1:1) conversations.
Figure~\ref{fig:client-fanout} illustrates that the sending client encrypts the same plaintext message separately for each group participant using the respective session keys.
Each encrypted message, or \textit{envelope}, is uploaded to the server, which relays it to the intended recipients without knowing its content. %
This approach preserves strict \ac{E2EE} semantics but incurs significant computational and bandwidth overhead for the sender in large groups, as the client must perform multiple encryptions and uploads per message.

\begin{figure}[ht]
    \centering
    \begin{subfigure}[b]{0.35\linewidth}
        \centering
        \includegraphics[width=\linewidth,page=1]{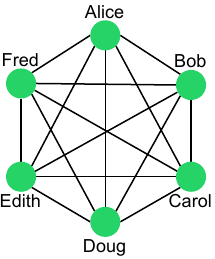}
        \caption{Pairwise E2EE}
        \label{fig:pairwise-cost}
    \end{subfigure}
    \hspace{5ex}
    \begin{subfigure}[b]{0.35\linewidth}
        \centering
        \includegraphics[width=\linewidth,page=2]{figures/SenderKey.drawio.pdf}
        \caption{Sender Key}
        \label{fig:senderkey-cost}
    \end{subfigure}
    \caption{Costs of different group messaging schemes.}
    \label{fig:cost-comparison}
\end{figure}

\subsection{Sender Key Protocol (Server Fan-Out)}
The \textit{Sender Key Protocol}~\cite{balbas_whatsupp_2023, balbas_analysis_2023, whatsapp_whatsapp_2023} optimizes the message complexity of group messaging by introducing a single (ephemeral) shared symmetric key for all group members.
Each participant independently chooses such a key and distributes it to all other members once, typically when sending their first message, using the same pairwise \ac{E2EE} channels as in client fan-out. 
Once the shared key has been distributed, the sender encrypts each message once using their individual Sender Key and uploads a single (signed) ciphertext to the server, rather than performing multiple encryptions.
The server then distributes this ciphertext to all group members via a broadcast mechanism (i.e., \textit{server fan-out}), as illustrated in Figure~\ref{fig:server-fanout}.
In large groups, this design reduces the sender's workload and message complexity, as well as network usage, shifting scalability concerns to the server, while maintaining confidentiality and authenticity through the key distribution mechanism built atop the Signal Protocol.

\paragraph{Current State.}
Both WhatsApp and Signal are based on the Signal protocol.
According to the WhatsApp Security Whitepaper~\cite{whatsapp_whatsapp_2023}, WhatsApp has employed the Sender Key protocol for group messaging since the rollout of E2EE in 2016.
In contrast, Signal relied on the client fan-out approach until August 2021, when it also transitioned to the more efficient Sender Key variant~\cite{signalandroid_sender_key}.
In late 2020, Signal additionally introduced the new private group system (also referred to as \emph{groups V2} or \emph{zkgroups})~\cite{chase_signal_2020,jimio_technology_2019,signal_messenger_new_2020},
which improves privacy by concealing group membership and other identifying information from the service provider.

In contrast, the other services investigated in this work (iMessage, Threema) currently use a pairwise client fan-out approach for \ac{E2EE} group communication.

\paragraph{Cost Comparison.}
In client fan-out, each message is encrypted separately for every recipient, resulting in $O(n)$ encryption operations and uploads per message.
Server fan-out reduces this to $O(1)$ by encrypting the message once and relying on the server for distribution.
This improvement applies to all ongoing group messages. While the initial distribution of the Sender Key relies on pairwise \ac{E2EE}\ channels and incurs a one-time cost, this overhead is outweighed by the efficiency gains of subsequent group messages.

Note that the examples in Figure~\ref{fig:cost-comparison} assume that each participant owns only a single device.
In practice, users often have multiple linked companion devices (e.g., tablets, laptops, or desktops).
Under the client fan-out model with per-device sessions, the sender must encrypt and upload a distinct ciphertext for every device belonging to every participant, further amplifying computational and bandwidth costs and -- as we show in this work -- potential \ac{TC} issues.
An exception is Threema, which relies on a single account-wide key that it shares across devices in multi-device settings~\cite{threema_threema_2025}.

\subsection{Transcript Consistency}

Properties offering some form of \emph{transcript consistency} (TC) were discussed under many names, e.g., \emph{consensus}~\cite{goldberg_multi-party_2009}, \emph{speaker consistency} and \emph{global transcript}~\cite{unger_sok_2015}, \emph{causality preservation}~\cite{unger_sok_2015,chen_integrating_2024}, \emph{transcript agreement}~\cite{cohn-gordon_ends--ends_2018}, \emph{transcript consistency}~\cite{marlinspike_private_2014,kleppmann_secure_2018}.
We group different terms and flavors into four TC variants. These variants range from the weakest to the strongest consistency guarantees for a message transcript. Note that we intentionally do not specify when or how often TC has to be checked. 
As none of the analyzed E2EE messaging services provides any form of TC, we defer a detailed contextualization and discussion to the Appendix~\ref{sec:TC}:
\begin{itemize}\parskip=3pt plus 3pt\itemsep=0pt
    \item \emph{set transcript consistency} (STC): The participants of the group eventually agree on the same (unordered) set of messages belonging to the conversation. 
    \item \emph{participant transcript consistency} (PTC): The participants of the group eventually agree on the ordered sequence of messages coming from each participant individually. Messages from different participants are not relatable to each other, except that they belong to the same conversation.  
    \item \emph{causal transcript consistency} (CTC): The participants of the group eventually agree on the same (unordered) set of messages belonging to the conversation, as well as a strict partial order of messages, i.e., not every pair of messages must be comparable (e.g., messages can be at the same height). 
    \item \emph{total transcript consistency} (TTC): The participants of the group eventually agree on the same ordered set of messages belonging to the conversation, s.t. for any two messages $m_1,m_2$ that appear in the transcript of an honest participant, if $m_1$ precedes $m_2$ for some honest participant, then $m_1$ precedes $m_2$ for every honest participant.
\end{itemize}

\section{Threat Model}
\label{sec:threat-model} 
We assume an asynchronous communication model, which is common in instant messaging environments~\cite{signal_technology_foundation__signalapp_libsignal-protocol-java_2022,marlinspike_private_2014} and distributed systems literature~\cite{cachin_introduction_2011}, i.e., there is no guarantee that messages are delivered in order, or within a known upper time bound.
In all analyzed instant messengers, all messages are delivered via a central server infrastructure operated by the respective instant messaging service. In other words, all messages must pass through a central server.
The trust that is put into the server varies across the analyzed services. While Signal aims to minimize the information available at the server~\cite{jimio_technology_2019,chase_signal_2020}, WhatsApp servers hold plain text information regarding group memberships\footnote{We revisit this aspect in more detail when discussing possible mitigation strategies in \Cref{sec:mitigations}.}.

\begin{table}[t]
    \centering
    \small
    \setlength{\tabcolsep}{1.2pt}
    \begin{tabularx}{\linewidth}{Xlrrp{2.48cm}}
        \toprule
        \multicolumn{1}{l}{\textbf{}} &
        \multicolumn{1}{l}{\textbf{Group}} &
        \multicolumn{1}{l}{\textbf{Max.}} &
        \multicolumn{1}{l}{\textbf{User}} &
        \multicolumn{1}{l}{\textbf{Open-Source}} \\
        \multicolumn{1}{l}{\textbf{Service}} &
        \multicolumn{1}{l}{\textbf{Messaging}} &
        \multicolumn{1}{l}{\textbf{Size}} &
        \multicolumn{1}{l}{\textbf{Estimate}} &
        \multicolumn{1}{l}{\textbf{Implementation}} \\
        \midrule
        WhatsApp~\iWhatsApp & Sender Key    & 1,024 & 3.5\,B & whatsmeow~\cite{whatsmeow}       \\
        iMessage~\iMessage  & Pairwise E2EE & 32    & 1.3\,B             & rustpush~\cite{rustpush}        \\
        Signal~\iSignal     & Sender Key    & 1,001 & 70\,M                & Signal-Desktop~\cite{signal_desktop}  \\
        Threema~\iThreema   & Pairwise E2EE & 256  & 12\,M               & threema-android~\cite{threema_android} \\
        \bottomrule
    \end{tabularx}
    \caption{Overview of the analyzed messaging services, their group messaging protocols, maximum group sizes~\cite{whatsapp_group_size,iMessage_group_size,signal_support_group_chats,threema_group_size,threema_ios_info_plist_2025,signalandroid_remoteconfig_2026}, estimated user bases~\cite{gegenhuber_2025_heythere,mau_imessage,mau_signal,mau_threema}, and the open-source implementations we used as bases.}
    \label{tab:messenger-selection}
\end{table}

\paragraph{Adversary.}
The adversary is a registered user of the messaging system and one participant of the targeted group chat\footnote{All described attacks work with at least one adversarial participant within a group.}.

The attacker fully controls their own devices and local protocol state, and thus may arbitrarily deviate from the prescribed protocol rules. 
The adversary is not capable of violating the security assumptions of cryptographic primitives, does not control or collude with the server, and does not have access to other users' long-term secrets or devices.
In practice, especially in large groups, participants can be added via invites or links and may not be uniformly trusted, which makes malicious or compromised members a realistic threat.

\paragraph{Adversarial Goals.}
The adversary aims to violate \ac{TC} within a group chat while preserving \ac{E2EE} semantics.
In particular, the adversary seeks to achieve the following without being detected by the server or through alerts or visible clues in the standard user interfaces offered for the respective chat software:

\begin{enumerate}\parskip=3pt plus 3pt\itemsep=0pt
    \item[\roundframe{G1}] \textbf{Inconsistent delivery to participants:}
    Cause different participants in a group conversation to receive different messages, or selectively prevent some participants from receiving a message, so that honest users observe inconsistent views of the group transcript.

    \item[\roundframe{G2}] \textbf{Inconsistent delivery to devices:}
    Deliver different messages to different devices belonging to the same honest user in a direct or group conversation, or drop messages for specific devices, causing the user’s devices to diverge in their view of the conversation history.
\end{enumerate}

The motivation for such attacks ranges from low-impact pranks to high-impact scenarios such as targeted device exploitation (\roundframe{G2}), as well as the deliberate disruption of group coordination by inducing disagreement and confusion (\roundframe{G1}), which is particularly consequential in large groups and high-stakes decision-making contexts.

\section{Methodology}
\label{sec:methodology}

In this section, we describe our messenger selection, experimental setup, and evaluation methodology.
In addition to analyzing the message transmission paths used for group communication (e.g., pairwise E2EE and the Sender Key protocol), we examine the robustness of each service against anomalies during key distribution and analyze the recovery and retransmission behavior triggered by encryption failures.
While this section investigates the general equivocation vectors and protocol weaknesses, Section~\ref{sec:results} evaluates exploitation in practice and demonstrates that an attacker can exploit these vectors not only for standard text messages but also for rich content such as location sharing or polls.

\subsection{Messenger Selection}\label{subsec:messenger_selection}
We aim to include relevant instant messaging services that support \ac{E2EE} in different variants (e.g., group messaging model, their target user bases, and free vs. paid usage).
The Signal protocol is widely regarded as the de facto gold standard for secure \ac{E2EE} messaging, and the Signal application is among the most popular messengers within the security community.
In addition, WhatsApp has approximately 3.5 billion active users~\cite{gegenhuber_2025_heythere} and adopted the Signal protocol in 2016.
We thus start our evaluation with these two applications and Signal's Sender Key protocol.

Beyond the Signal protocol, other instant messengers have adopted E2EE as well.
For example, iMessage is preinstalled on iPhones and other Apple devices, is among the most popular instant messengers in the US, and accounts for more than one billion users globally.
Finally, we include Threema as an additional messenger with a strong focus on security and privacy; it is the only paid app among the services we consider and is particularly popular in Europe.

In \Cref{tab:messenger-selection}, we provide an overview of the messenger applications we study, along with their estimated user bases and current maximum group sizes.

\begin{figure}[t]
    \centering
    \includegraphics[width=\linewidth]{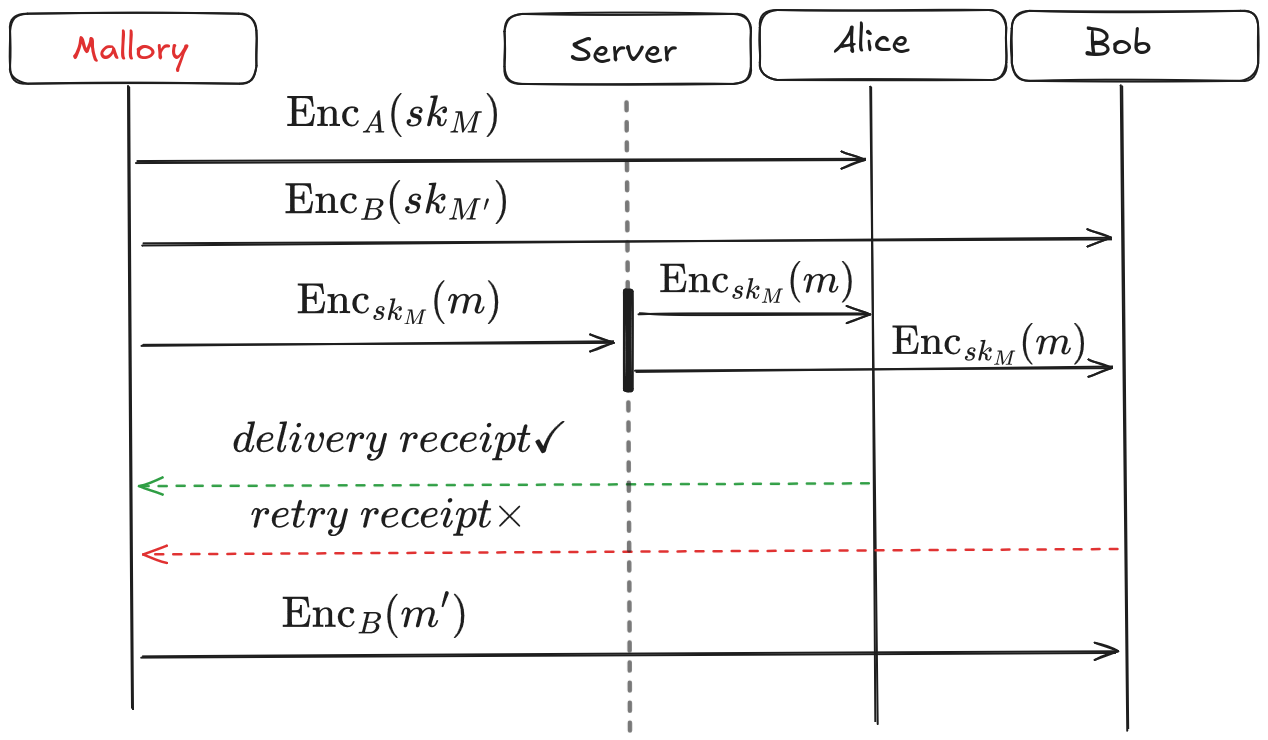}
    \caption{Group communication with sender key messages (WhatsApp, Signal), where a decryption failure triggers retransmission of a message via direct \ac{E2EE} communication.}

    \label{fig:case_1}
\end{figure}

\subsection{Experimental Setup and Evaluation}

To reflect the adversarial capabilities described in Section~\ref{sec:threat-model} (i.e., control over device state, their encryption keys, and the ability to deviate from the protocol, while still sending syntactically valid messages), we use both official reference clients (Signal, Threema) and unofficial client implementations (WhatsApp, iMessage). 
We apply targeted patches that enable controlled deviations from the protocol behavior.
While open or unofficial clients simplify testing, an attacker could also reverse engineer the official clients and perform the experiments via instrumentation, e.g., using Frida~\cite{frida}.
We assume the attacker to be an existing member of the group, while one or more other group participants act as victims.
On the victim side, we evaluate clients across major platforms, including Android, iOS, Windows, macOS, Linux, and web-based clients.

To assess the robustness of the target applications with respect to message delivery and \ac{TC}, we evaluate their behavior under a range of error conditions and atypical delivery paths, including cases where protocol assumptions are violated.
For each messenger, we first analyze the default group message flow.
For WhatsApp and Signal, we start with Sender Key-based group messaging and then evaluate fallback behavior triggered by decryption errors. %
Next, we examine pairwise \ac{E2EE} transmission, which is the default for iMessage and Threema, and test whether similar pairwise delivery can also be used in WhatsApp and Signal to send group messages.
While the previous evaluations focus on equivocation via pairwise \ac{E2EE} channels, we finally assess how robust Sender Key-based message broadcast (i.e., on WhatsApp and Signal) is against equivocation and whether contradictory or repeated messages via the broadcast channel trigger user-visible alerts.

\begin{figure}[t]
    \centering
    \includegraphics[width=\linewidth]{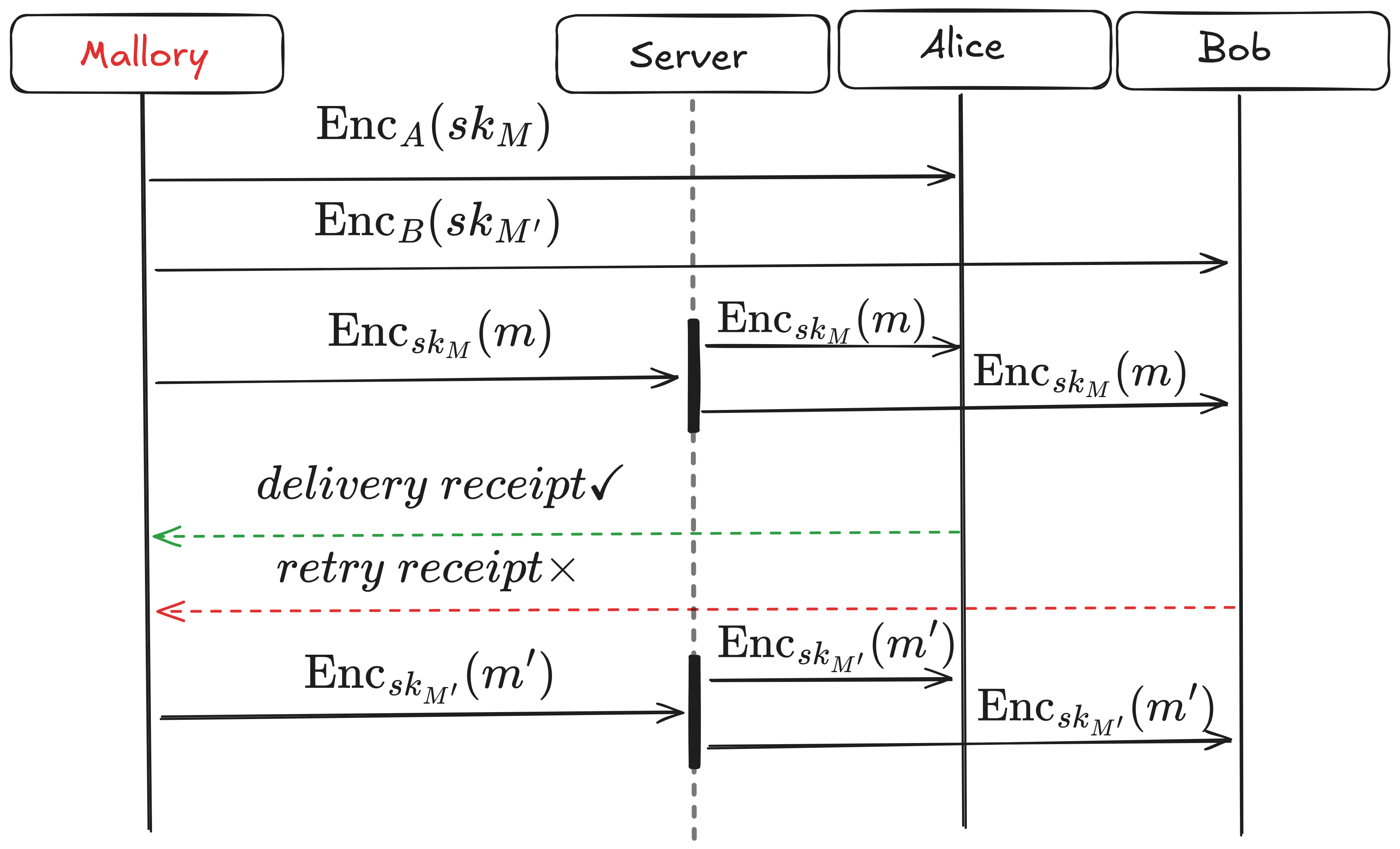}
    \caption{Message equivocation using sender key sends only (WhatsApp, Signal). We test whether multiple broadcasts that result in contradictory or repeated message deliveries constitute a practical attack vector, or whether they trigger UI warnings or suspicion at receiving clients.}
    \label{fig:case_3}
\end{figure}

\subsubsection{Sender Key with Pairwise E2EE Fallback}
Both WhatsApp and Signal employ the Sender Key protocol for group messaging.
Under this protocol, each group message is encrypted using a symmetric \emph{sender key} selected by the message sender.
To enable other participants to decrypt future Sender Key messages, the sender first distributes the sender key to all group members via pairwise E2EE channels.

Figure~\ref{fig:case_1} demonstrates that this design permits a malicious sender to distribute different sender keys to participants, which would result in decryption failures for one or more recipients.
We investigate the resulting failure and recovery behavior and show that, in practice, failures of the Sender Key protocol are often handled by falling back to pairwise E2EE communication.
While this behavior is the intended recovery path, the resulting pairwise sessions can subsequently be used to transmit different messages to different participants.

\begin{figure}[t]
    \centering
    \includegraphics[width=\linewidth]{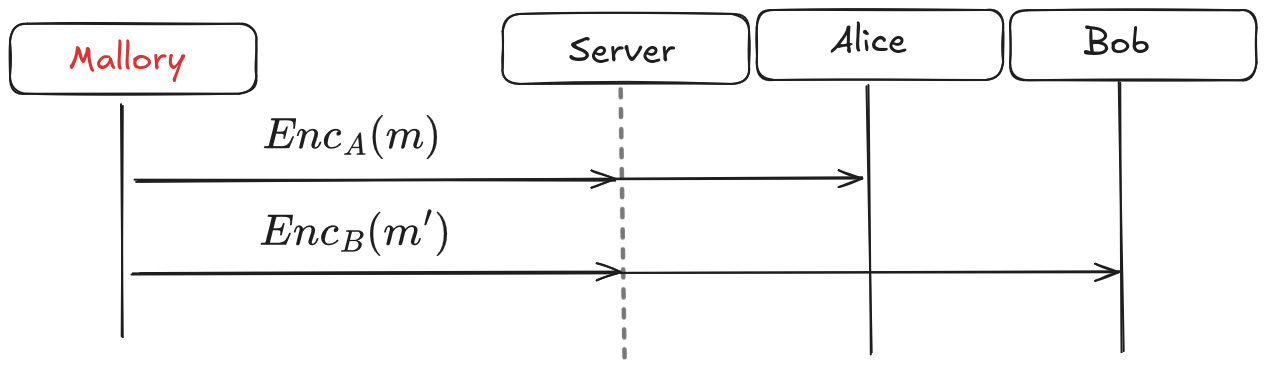}
    \caption{Group communication based on pairwise \ac{E2EE} is inherently prone to equivocation attacks. While some messengers still use it as the default transmission path in groups (iMessage, Threema), we evaluate whether it can nevertheless be leveraged for group communication on the messengers supporting the Sender Key protocol (WhatsApp, Signal).}
    \label{fig:case_2}
\end{figure}

\subsubsection{Pairwise E2EE Messages}
For iMessage and Threema, this is the regular (and only) path for group communication, being highly vulnerable to message equivocation by design, as demonstrated in Figure~\ref{fig:case_2}.

Signal adopted the Sender Key protocol for group messaging in 2021 and relied on pairwise E2EE group messages for much of its lifetime.
Both Signal and WhatsApp employ pairwise E2EE transmission as a recovery mechanism for group messages when Sender Key sends fail.
Because encryption failures are sometimes visible to users in the client UI, we further investigate whether a malicious client can intentionally bypass the Sender Key protocol and instead transmit group messages exclusively via the pairwise E2EE path as the primary delivery mechanism.

\subsubsection{Sender Key Messages Only}
The previous two techniques demonstrate how a malicious sender can create divergent views of the message transcript by deviating from the Sender Key protocol (if applicable) and using directed, pairwise \ac{E2EE} channels to transmit equivocated messages.
In this final scenario, shown in Figure~\ref{fig:case_3}, we examine whether an attacker can violate \ac{TC} while using only Sender Key messages, without relying on transmitting the equivocated messages via pairwise \ac{E2EE} sessions.
Specifically, we test how clients respond when an attacker broadcasts multiple versions of the same message, each encrypted for different clients, using the server fan-out mechanism.

\section{Results and Exploitation}
\label{sec:results}
In this section, we first present transcript inconsistencies under the analyzed protocols. We then discuss practical attack scenarios and highlight implementation-specific issues.

\subsection{Equivocation Vectors}
Table~\ref{tab:vulnerable-service-overview} shows that all services are vulnerable to message equivocation in the group setting (\roundframe{G1}).
Threema is not vulnerable to the device-specific equivocation case (\roundframe{G2}) because of its distinctive multi-device architecture: only a single device of the receiving account fetches the message from the server, then re-encrypts and reflects it to all remaining devices via a mediator server.
This is a consequence of Threema's multi-device implementation rather than stronger \ac{TC} defenses, and it comes with a serious trade-off, since Threema's multi-device protocol implementation does not support forward secrecy~\cite{threema_threema_2025}.

\begin{table}
\setlength{\tabcolsep}{7pt}
    \centering
    \newcommand\tWhatsApp{\tikz[baseline=(O.base)]{\node(O) [baseline,minimum width=3mm,inner sep = 0,align=left] {\iWhatsApp};}}
    \newcommand\tSignal{\tikz[baseline=(O.base)]{\node(O) [baseline,minimum width=3mm,inner sep = 0,align=left] {\iSignal};}}
    \newcommand\tIMessage{\tikz[baseline=(O.base)]{\node(O) [baseline,minimum width=3mm,inner sep = 0,align=left] {\iMessage};}}
    \newcommand\tThreema{\tikz[baseline=(O.base)]{\node(O) [baseline,minimum width=3mm,inner sep = 0,align=left] {\iThreema};}}
    \begin{tabularx}{\linewidth}{X cccc p{0.1mm} cccc}
        \toprule
        \textbf{}
            & \multicolumn{4}{c}{\textbf{One-to-One (1:1)}} &
            & \multicolumn{4}{c}{\textbf{Group Chats}} \\ \cmidrule(lr){2-5} \cmidrule(lr){7-10}
        &
            \tWhatsApp
            & \tIMessage
            & \tSignal
            & \tThreema
            &
            & \tWhatsApp
            & \tIMessage
            & \tSignal
            & \tThreema \\
        \midrule
        \roundframe{G1}
            & \neutralMinus & \neutralMinus         & \neutralMinus         & \neutralMinus    &    
            & \greenCheck & \greenCheck & \greenCheck & \greenCheck \\
        \roundframe{G2}
            & \greenCheck & \greenCheck & \greenCheck & \redRemove &
            & \greenCheck & \greenCheck & \greenCheck & \redRemove \\
        \bottomrule
    \end{tabularx}
    \\
    \vspace{1ex}
    \roundframe{G1} Participant inconsistency \hspace{1.5ex}
    \roundframe{G2} Device inconsistency\\
    \neutralMinus not applicable  \hspace{1.5ex}
    \greenCheck vulnerable  \hspace{1.5ex}
    \redRemove not vulnerable
\caption{Overview of messengers vulnerable to \roundframe{G1}, which enables participant inconsistency, and \roundframe{G2}, which enables device inconsistency, in 1:1 communication and group chats.}
\label{tab:vulnerable-service-overview}
\end{table}

\subsubsection{Sender Key with Pairwise E2EE Fallback}
\begin{figure}[t]
  \centering
  \includegraphics[width=0.9\linewidth]{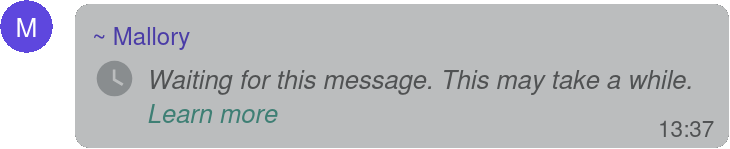}
    \caption{By default, WhatsApp shows a decryption error message bubble upon unsuccessful message decryption. However, the error persists only until the sender retransmits the message. Moreover, WhatsApp lets the sender suppress these warnings on the receiver side, again leading to silent failure and leaving no trace of the attack in the victim's UI.}
      \label{fig:whatsapp-decrypt-fail}
\end{figure}

Both WhatsApp and Signal implement a recovery mechanism for Sender Key group messages in which recipients can request a retransmission after a decryption failure. 
In response, the sender retransmits the message to the requesting recipient via the pairwise 1:1 \ac{E2EE} session, which provides an opportunity for equivocation by delivering recipient-specific content (cf. Figure~\ref{fig:case_1}).
In Signal, unsuccessful message decryptions fail silently, thus creating no barriers for the attacker.
In WhatsApp, decryption errors are shown to the user, as demonstrated in Figure~\ref{fig:whatsapp-decrypt-fail}.
While this could potentially raise suspicion by the user and thereby limit exploitability, the error bubble disappears immediately after 
the attacker retransmits the equivocated message, thus only being visible for a fraction of a second.
More severely, WhatsApp allows the message sender to set a \emph{decrypt-fail} attribute to \emph{hide}, effectively suppressing this error bubble entirely on the receiver side, yielding the same behavior and exploitability as in Signal.
To summarize, this vector enables sending different messages to arbitrary participants and their devices, achieving both goals \roundframe{G1} and \roundframe{G2} in WhatsApp and Signal.

\subsubsection{Pairwise E2EE Messages}
While Sender Key messages aim to construct a shared broadcast channel, messaging based on pairwise \ac{E2EE} sessions is inherently vulnerable to equivocation.
We confirm this vector for iMessage and Threema, both of which exclusively rely on pairwise channels for group communication.

Although WhatsApp and Signal could, in principle, enforce Sender Key delivery for all group messages except for sender key distributions, both messengers still accept group messages sent via pairwise channels. 
In Signal, this requires no special effort, as messages can simply be sent via the legacy group messaging path used prior to 2021 (cf. Section~\ref{sec:background}). 
In WhatsApp, pairwise group messages are rejected by default, but we found that they are accepted when disguised as retransmissions. 
This holds even if the receiving client never requested a retransmission, effectively enabling arbitrary equivocation via pairwise \ac{E2EE} channels. 
In summary, this equivocation vector is viable across all studied messengers and does not rely on decryption failures that could otherwise serve as a detection signal. 
In addition to \roundframe{G1}, if the protocol uses per-device sessions, which is the case for all evaluated services except Threema, targeting specific devices (\roundframe{G2}) is also possible.

\subsubsection{Sender Key Messages Only}
Instead of responding to a victim's retry request with a pairwise retransmission, both WhatsApp and Signal allow a malicious sender to ignore the retry receipt and simply broadcast the equivocated message via the Sender Key approach.
Clients who have already successfully received the corresponding message (matched by its unique message ID) on the first attempt simply ignore the subsequent transmission, even if it would cause a decryption failure or contain contradicting content.
Again, no warnings were displayed to the user.

On WhatsApp, the server tracks all members of a group and always broadcasts a message to all participants.
To improve privacy, Signal does not store this kind of data in plaintext and thus allows the message sender to define who is part of the group and should be included in the message broadcast.
This creates an additional vector for creating individual transmissions for different group participants, effectively allowing the attacker to establish individual channels to each participant using Sender Key group messages\footnote{This differs from pairwise delivery because 1:1 messages and Sender Key group messages are generally handled via different server endpoints. Additionally, the Signal server even allows Sender Key messages (which are multi-recipient sealed sender V2 messages) through the 1:1 endpoint~\cite{signalserver_acceptedtypes_2025}.}.
To summarize, via multiple Sender Key message broadcasts, the attacker is able to achieve both goals (\roundframe{G1}, \roundframe{G2}) for all evaluated and applicable applications.

\begin{whatsapptakeaway}
We show that:
\begin{itemize}[nosep,leftmargin=1.5em]
    \item[\deviceAck] pairwise \ac{E2EE} messaging is vulnerable to transcript inconsistency attacks per design;
    \item[\deviceAck] current Sender Key implementations are vulnerable due to fallback and retransmission paths;
    \item[\deviceAck] thereby, all evaluated messenger services are vulnerable to message equivocation, and none provide server-side or UI detection mechanisms, enabling practical, stealthy exploitation.
\end{itemize}
\end{whatsapptakeaway}

\subsection{Exploitation Examples}
After demonstrating the possibility of sending inconsistent messages within a group, we show the impact of the previously described equivocation
vectors by providing examples that could occur in real-world scenarios.
While we earlier showed that all studied messengers are vulnerable to transcript inconsistencies, polls are particularly critical because their results are rendered as authoritative application state.
In addition to design-level vectors, we therefore analyze implementation-specific behaviors and bugs.
Although not required to violate \ac{TC}, these issues are practically relevant because they amplify \ac{TC} failures by reducing attacker effort and suppressing UI warnings. %
Alongside this paper, we release videos demonstrating these attacks against unmodified, real-world client applications and devices for all investigated messengers\footnote{Attacks were executed exclusively against our own accounts and devices.}.

\begin{figure}[t]
    \centering
    \begin{subfigure}[b]{0.39\linewidth}
    \centering
    \includegraphics[width=\linewidth]{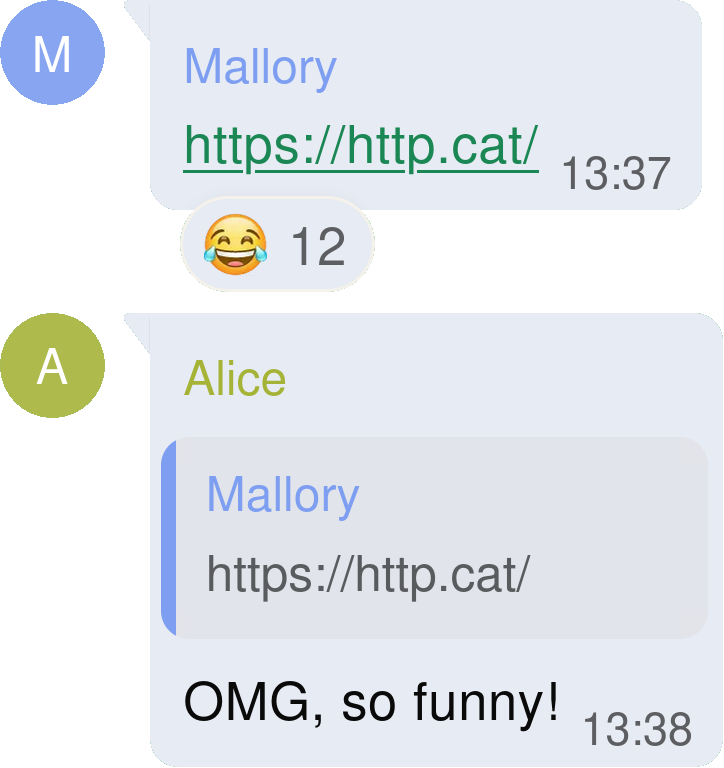}
    \caption{Normal member view.}
    \end{subfigure}
    \hspace{2ex}
    \begin{subfigure}[b]{0.39\linewidth}
    \centering
    \includegraphics[width=\linewidth]{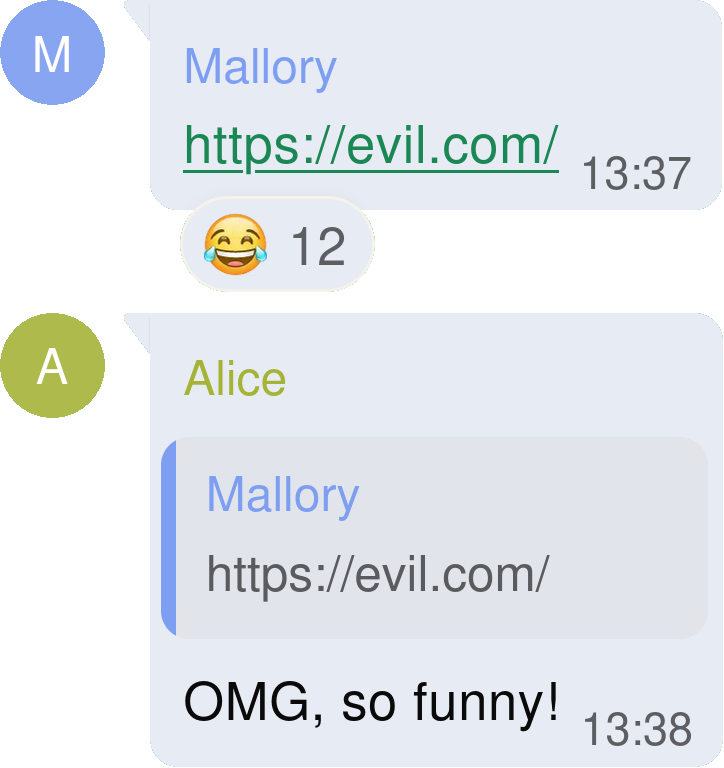}
    \caption{Victim's view.}
    \end{subfigure}
    \caption{Split-view attack in which the attacker (Mallory) sends a benign link to most group members and exploits their reactions to increase trust in a spoofed link sent to the victim.}
    \label{fig:whatsapp-url-trust}
\end{figure}

\subsubsection{Social Engineering and Manipulation Tactics}
Attackers can exploit inconsistencies in group chats to manipulate group members through social engineering and deliberate deception.
\Cref{fig:whatsapp-side-by-side} illustrates how a malicious group member can sabotage group coordination in the context of a scheduled meeting.
Victims may receive no message at all, or receive entirely different meeting details.
The technique is not limited to text-based messages; for example, on WhatsApp, attackers can apply it to the location-sharing functionality.
To further increase deniability, attackers can leverage short message timers to automatically delete messages, removing evidence and enabling plausible deniability after the incident.
In addition, such manipulation can undermine the victim’s trust in their own perception and memory, which may cause psychological distress and negatively affect mental well-being.

\subsubsection{Evading Moderation and Split View Attacks}
Attackers could also exploit transcript inconsistencies for phishing campaigns and to evade moderation.
Similar to \Cref{fig:whatsapp-side-by-side}, an attacker can send messages with different content to group participants, for example, benign-looking messages to group administrators and moderators while simultaneously delivering scam messages to other group members.

A related scenario combines phishing links with legitimate URLs.
Since replies and reactions (e.g., emoji) are linked to messages by their message ID, an attacker can induce part of the group to react positively (e.g., with \eThumbup) to a benign URL, such as a cat picture.
The same reaction can then appear attached to the attacker's spoofed message for members who instead received the malicious link, thereby increasing its perceived trustworthiness. An example of this attack is demonstrated in Figure~\ref{fig:whatsapp-url-trust}.

\begin{figure}[t]
    \centering
    \includegraphics[width=0.9\linewidth]{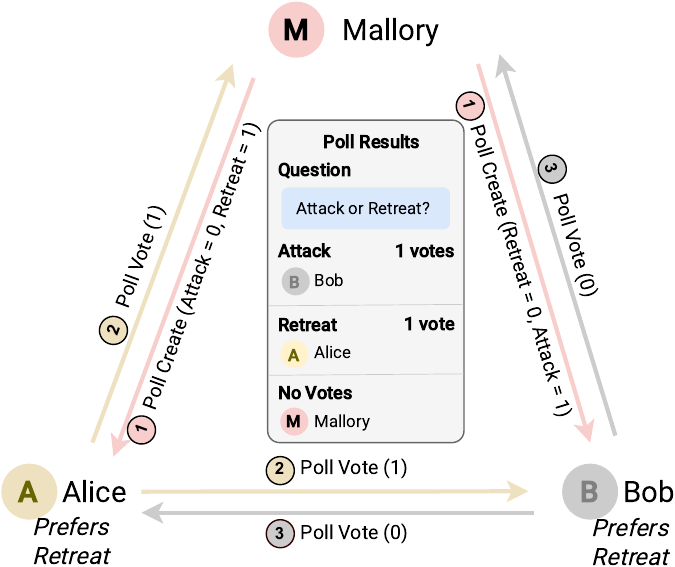}
    \caption{Mallory sends a manipulated poll to Bob (\ding{202}). For Mallory and Alice, the option \emph{Attack} has index 0 and \emph{Retreat} has index 1. For Bob, the indices are swapped. Although Bob, like Alice (\ding{203}), prefers \emph{Retreat} (\ding{204}), his vote appears as \emph{Attack} to the other participants. Due to the tie in votes, Mallory can now decide whether to attack or retreat.}
    \label{fig:changed_options}
\end{figure}

\subsubsection{Rigging Polls}
While all investigated messaging services support polls today, the feature was introduced at different times.
Threema was the first to introduce them in January 2015~\cite{threema_polls}.
WhatsApp introduced polls for group chats in November 2022~\cite{whatsapp_2022_polls} and refined the feature in May 2023~\cite{whatsapp_2023_polls}.
More recently, iMessage added polls in June 2025 with iOS~26~\cite{imessage_polls}, and Signal introduced the feature in November 2025~\cite{signal_polls}.

We begin with an analysis of how polls are implemented across different messaging services.
We identified three distinct poll action types:
\begin{itemize}\parskip=3pt plus 3pt\itemsep=0pt
    \item \textbf{Poll Create:} A group member creates the poll by specifying the question and available options.
    \item \textbf{Poll Vote:} Each participant casts a vote for one or more options.
    \item \textbf{Poll Close:} In Signal and Threema, the poll creator can finalize the votes and close the poll.
\end{itemize}

Across all services, subsequent poll actions (i.e., \textit{voting} and \textit{closing}) reference the original \textit{poll creation} message via its message ID.
The \textit{voting} action is implemented differently across services: WhatsApp encodes votes using a hash of the selected option, whereas Signal and Threema use an index, and iMessage relies on a UUID-based identifier.

In the following, we consider two types of attackers: malicious poll creators and voters.

\begin{figure}[t]
    \centering
    \begin{subfigure}[b]{0.41\linewidth}
    \centering
    \includegraphics[width=0.85\linewidth]{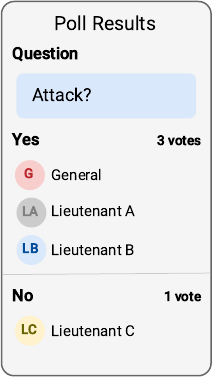}
    \caption{Members were lured into voting for the attack option.}
    \end{subfigure}
    \hspace{2ex}
    \begin{subfigure}[b]{0.41\linewidth}
    \centering
    \includegraphics[width=0.85\linewidth]{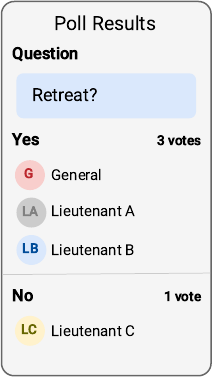}
    \caption{View of victims receiving an inverted question.}
    \end{subfigure}
    \caption{A malicious poll creator can send different questions to parts of the group, tricking them into voting for unfavored options and leading to manipulated results.
    Attacking polls by sending out inverted questions is also possible on WhatsApp, since voting hashes only consider the voting option text. We provided videos of this attack in our artifact.}
    \label{fig:signal-poll-changed-question}
\end{figure}

\paragraph{Malicious Poll Creator.}

A malicious poll creator can send different poll creation messages to different participants, leading to divergent local interpretations.

\smallskip
\noindent
\emph{Juggling Poll Options.}
For index- or ID-based schemes (Signal, Threema, iMessage), the attacker can reorder poll options or provide entirely different questions or option sets across recipients as demonstrated in Figure~\ref{fig:changed_options}.

In hash-based schemes (i.e., WhatsApp), the attacker can manipulate poll options by introducing insignificant changes (e.g., whitespaces or encoding variations), thereby producing unrecognized hashes.
Votes cast for these manipulated options are silently discarded without error for other participants.

\smallskip
\noindent
\emph{Inverting Poll Questions.}
Across all services, the poll question itself is not protected by cryptography. Thus, an attacker can change it arbitrarily. For binary questions, this can be abused, even when hash-based voting is used (i.e., on WhatsApp), by inverting the question as shown in Figure~\ref{fig:signal-poll-changed-question}.
If hashing covered not only the selected voting option but also the corresponding poll question, such attacks would not be possible.

\smallskip
\noindent
\emph{Dictating and Orchestrating Poll Results.} 
In Threema, the poll creator rebroadcasts all poll votes previously collected by other participants for intermediate results.
Thus, a malicious poll creator can add, remove, or rewrite votes for any group participant upon closing the poll, thereby altering the final result. The creator can close the poll multiple times in Threema, overwriting the entire result repeatedly (including the title/description and the options).

Signal does not require the creator to broadcast the final poll state upon closing. Instead, it sends a message to participating clients to close the poll on their end, referencing the original poll message by timestamp ID. The clients ignore future votes referencing an already closed poll. Thus, selectively sent poll close messages could make some members think a poll is still ongoing, when other members already see a final result.

\begin{figure}[t]
  \centering
  \hspace{2ex}
  \begin{subfigure}[b]{0.25\linewidth}
    \includegraphics[width=1\linewidth]{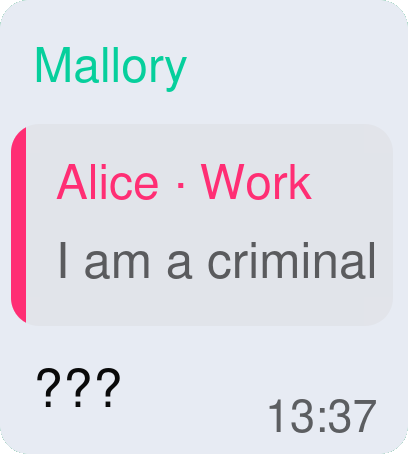}
    \caption{WhatsApp}
    \label{fig:whatsapp_fake_quote}
  \end{subfigure}
  \hfill
  \begin{subfigure}[b]{0.47\linewidth}
    \includegraphics[width=1\linewidth]{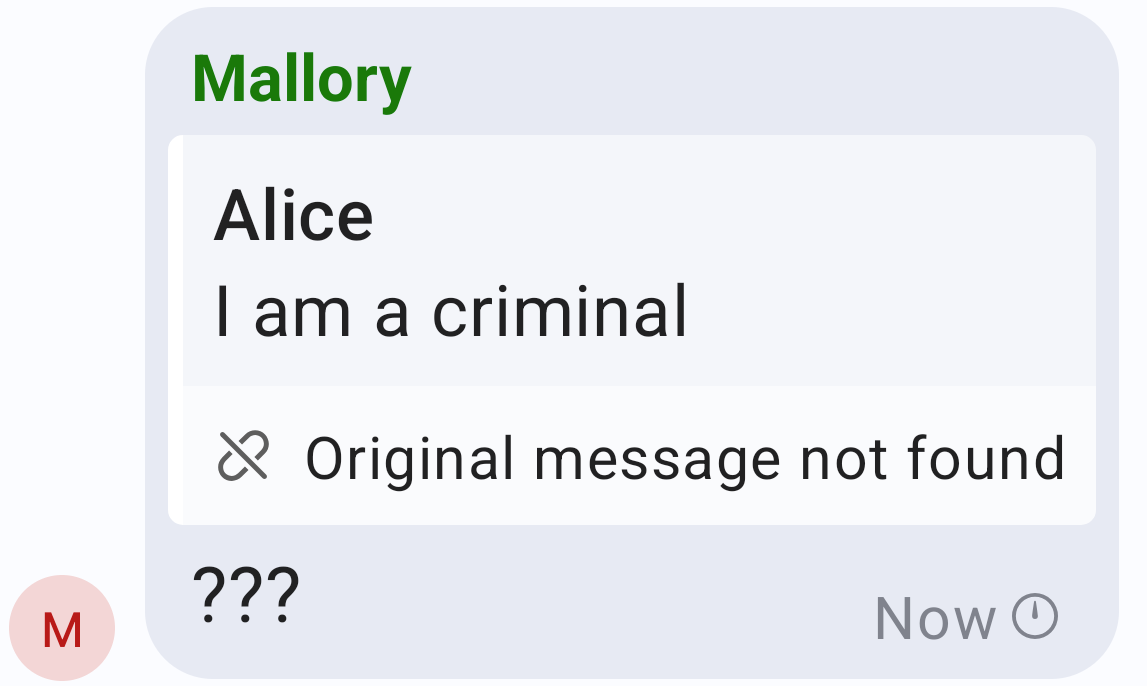}
    \caption{Signal}
    \label{fig:signal_fake_quote}
  \end{subfigure}
  \hspace{2ex}
  \caption{Screenshots of fake quote references on WhatsApp and Signal. A message references a nonexisting message, causing the client to display the attacker-provided fallback text. Signal shows that it does not find the referenced message.
  }
  \label{fig:fake_quote}
\end{figure}

\paragraph{Malicious Poll Voter.}
A malicious poll voter can equivocate by sending different voting messages to different participants.
As a result, intermediate poll results displayed on recipients’ devices may diverge, leading to inconsistent local views of the poll outcome. 

For Threema, the main attack vector consists of sending manipulated votes to the poll creator, as the poll creator’s final message overwrites intermediate results on other recipients’ devices.
Specifically, using a positive value different from 1 in a vote causes the vote to be counted toward the option’s total number of votes, while the UI does not indicate that the user selected the option.
As a result, a malicious voter can inject additional \emph{hidden votes}, even in single-choice polls, and send them to the creator to invalidate the final poll outcome.

\begin{whatsapptakeaway}
We show that transcript inconsistencies enable:
\begin{itemize}[nosep,leftmargin=1.5em]
\item[\deviceAck] conducting social engineering attacks, e.g., by sending different message content to different participants or by dropping messages;
\item[\deviceAck] evading moderation, e.g., by sending benign URLs to moderators and most participants, while delivering phishing links to selected victims;
\item[\deviceAck] manipulating polls, both as poll creator and as voter, due to transcript inconsistency, amplified by vulnerable design decisions, e.g., index-based voting.
\end{itemize}
\end{whatsapptakeaway}

\subsection{Implementation-Specific Issues} %
In addition to protocol-level observations, we identified several implementation-specific behaviors that introduce side channels and security risks.
We found potentially problematic behavior related to \ac{TC} in message quoting (WhatsApp, Signal) and implementation flaws that allow an adversary to submit votes on behalf of other participants (iMessage).
Moreover, we observed notable differences in the client behavior across operating systems and device types (especially in WhatsApp and Signal), which can leak metadata or enable device fingerprinting in adversarial settings.

\subsubsection{Quoted Messages}
We identified potential issues related to quoted messages. %
While Threema and iMessage require the quoted message (referenced by its message ID) to exist and do not include fallback information in a sent reply, Signal and WhatsApp embed fallback text that is rendered if the referenced message cannot be resolved (e.g., for recently joined group participants).
As a result, an adversary can craft messages that appear to quote content allegedly written by another participant, even though such a message was never sent, by providing a broken reference with attacker-controlled fallback text.

Figure~\ref{fig:fake_quote} illustrates this behavior in the official WhatsApp and Signal clients.
Although Signal's UI indicates that it cannot find the original message, an  attacker can potentially exploit this behavior in practice, e.g., when a newly added group member sees a fabricated quote and lacks the context required to assess its authenticity.
In contrast, WhatsApp does not display any warning alongside the quoted message, which directly facilitates social engineering attacks.

For Signal's desktop client~\cite{signal_desktop}, we additionally observed that the UI allows references to messages received after the quoting message, thereby violating causal ordering.
We provide a \ac{PoC} video in our artifact~\cite{signal_video_14}.

Additionally, we noticed implementation/OS-specific differences in resolving collisions within the message ID on Signal, see \Cref{sec:appendix:signal-inconsistent-duplicate-handling}.

\subsubsection{iMessage: Voting for Other Participants}

Apple implements polls as a protocol extension named \texttt{MSMessageExtensionBalloonPlugin}. Internally, the client represents a poll as \texttt{JSON}%
.
Also, for voting, the voter sends a \texttt{JSON} that includes a list of votes, each having a \texttt{voteOptionIdentifier}, the UUID of the option, and the \texttt{participantHandle}%
. 

The voter can manipulate the \texttt{participantHandle} field in the vote message, e.g., by changing it to another group member or by inserting additional votes attributed to different participants. In this case, the poll overview suggests that another member voted for an option or that multiple participants voted for it%
. However, the detailed poll view still correctly attributes the vote, since it counts it towards the sender and ignores the \texttt{participantHandle}.

\subsubsection{WhatsApp: OS-Specific Parsing of Polls}
We observed differences in how polls are parsed and interpreted between Android and iOS clients.
WhatsApp supports multiple user addressing formats, namely the \emph{JID} (Jabber ID) and the \emph{LID} (Logical ID), which decouples the user identifier from the phone number.
Official implementations use the newer LID format when hashing and encrypting votes.
However, we found that an attacker can craft votes using the legacy JID format, which are correctly parsed and displayed on Android but discarded and not shown on iOS devices.

Moreover, Android clients are more permissive when processing malformed \emph{poll creation} messages.
In particular, the Android client correctly parses and accepts polls containing repeated options, whereas iOS clients discard such polls entirely.

Conversely, during the voting phase, iOS correctly parses and accepts voting messages containing repeated (i.e., identical) hashes\footnote{The vote is nevertheless counted only once.}, whereas Android clients discard such messages.

In all three cases, an attacker can exploit these inconsistencies to induce divergent voting outcomes (e.g., by only letting Android users participate in the vote) without relying on the equivocation vectors described earlier.

Finally, we observed that the metadata appended to vote messages differs between Android and iOS, allowing an attacker to infer the sender's operating system by inspecting message metadata.

\begin{table}[t]
\centering
\newcommand\tFa{$^{\mathrm{a}}$}
\newcommand\tFb{$^{\mathrm{b}}$}
\begin{tabularx}{\linewidth}{l X l r}
\toprule
\textbf{OS} & \textbf{Client}          & \textbf{Prefix} & \textbf{Len} \\
\midrule
Android & Main Device                  & \texttt{A5}   & 32 \\ %
Android & Companion (e.g. Tablet)      & \texttt{AC}   & 32 \\ %
Android & Wear OS (e.g., SmartWatch)   & \texttt{A3}   & 32 \\
iOS     & Main, Companion, watchOS     & \texttt{3A}   & 20 \\
iPadOS  & Tablet iPadOS                & \texttt{3C}   & 20 \\
macOS   & macOS Desktop                & \texttt{3B}   & 20 \\
Windows & Windows Desktop UWP\tFa      & \texttt{3F}   & 20 \\
Windows & Windows Desktop Electron\tFb & \texttt{3EB0} & 22 \\
Web     & WhatsApp Web                 & \texttt{3EB0} & 22 \\
\bottomrule
\end{tabularx}
\\
\vspace{1ex} %
\tFa~Native UWP App was discontinued in Nov 2025.\\
\tFb~Current version uses Electron WebView~\cite{wa_win_webview}.
\caption{WhatsApp leaks a user's OS and device type through the prefix and length of generated message IDs.
}
\label{tab:wa-prefix-lengths}
\end{table}

\subsubsection{WhatsApp: OS-Specific Message IDs}
During our analysis across platforms, operating systems, and client configurations, we found that WhatsApp message IDs for arbitrary message types are not generated uniformly and can leak the sender's OS. 
Table~\ref{tab:wa-prefix-lengths} summarizes the observed prefix and length values for messages on different OSs and device types. 
In an adversarial setting, such leakage can facilitate targeted social engineering and aid reconnaissance by enabling an attacker to fingerprint a victim's devices and tailor subsequent attacks or exploits accordingly.
This privacy leak may also be abused in private relationships, for example, for stalking, by inferring that a message was sent from a specific device (e.g., a desktop computer), since the ID of a received message can be readily inspected in WhatsApp Web using standard browser developer tools by moderately skilled users.

\begin{whatsapptakeaway}
We show that the current implementations allow:

\begin{itemize}[nosep,leftmargin=1.5em]
\item[\deviceAck] putting words into a victim's mouth by fabricating quoted messages in WhatsApp and Signal;
\item[\deviceAck] spoofing votes for other participants for iMessage;
\item[\deviceAck] poll tampering by exploiting OS-specific implementation differences between WhatsApp clients;
\item[\deviceAck] fingerprinting the OS type of a message's sending client via the message ID in WhatsApp.
\end{itemize}

\end{whatsapptakeaway}

\section{Countermeasures and Mitigation Paths}\label{sec:mitigations}
Early academic papers already discussed solutions to check for \ac{TC} in retrospect~\cite{goldberg_multi-party_2009}, or proposed entirely new messaging protocol designs to address the issue of \ac{TC}~\cite{schliep_end--end_2019,schliep_consistent_2018}.
Apparently, none of those proposals were ever adopted for any of the major messaging services analyzed in this paper. 

We therefore take a different approach.
Instead of proposing another complete protocol (re-)design to address \ac{TC}, 
we analyze which form of \ac{TC} can be achieved within the currently deployed group messaging designs with as few modifications to protocol and threat model as possible.
Specifically, we propose minimal modifications to the Sender Key protocol used by two of the four analyzed prevalent instant messengers: WhatsApp and Signal.
Our changes enable messengers to issue user-facing warnings when \ac{TC} cannot be guaranteed, thereby making potential attacks transparent. In the best case, our modifications can ensure that all participants either know that STC might be violated, or that STC can eventually be achieved. %
This means that all group participants can arrive at a consistent transcript, given that they are online at the same time and new messages are submitted to the group slower than the network and processing delay of the slowest participant\footnote{See Appendix~\ref{sec:TC} for details.}.

More generally, we argue that improving \ac{TC} efficiently, i.e., without increasing communication complexity significantly, seems most realistic for group chats that operate in a common broadcast domain with shared group keys, where recipients can more readily detect inconsistencies and systems can provide meaningful user-facing warnings.
This is a strong argument for group chat applications with \ac{TC} to be built on top of MLS~\cite{barnes_messaging_2023,beurdouche_messaging_2025}.
Nevertheless, our findings highlight that explicitly considering \ac{TC} at the design level remains crucial for current and future group messaging systems.

\subsection{Sender Key with TC Improvements}
\label{sec:sender-key-tc-improvements}
To enhance the Sender Key protocol, we leverage the fact that all analyzed \ac{E2EE} messaging designs already rely on a centralized server infrastructure to relay messages.
If the server is assumed to be honest with respect to message delivery, similar to the trusted delivery service in MLS, addressing \ac{TC} becomes significantly easier.
We do not consider this assumption to constitute a substantial change to the trust or threat model, since the server already has the technical ability to arbitrarily drop messages and therefore must be trusted with respect to message delivery in any case.

Overall, our assumptions and rules for our improvements to Sender Key TC are: 
\begin{itemize}\parskip=3pt plus 3pt\itemsep=0pt
    \item \textbf{Unsolicited Retransmissions:} Client implementations must drop unsolicited 1:1 group retransmissions.
    
    \item \textbf{Enforced Group Broadcast:}
    It can be enforced that a message to a group conversation is sent to all its members.
    In other words, clients must only accept messages from the server transmitted via the Sender Key protocol to the entire group, 1:1 group messages are only allowed to distribute a sender key.%

    \item \textbf{Broadcast Retry Requests:}
    If a retry for a message in a group chat is requested by any participant, this retry request must be announced via server fan-out to all group participants, so that other honest members are informed about a message being dropped or retransmitted for another participant.

    \item \textbf{Static Group:} If a member joins or leaves, we treat this as a new group. Thus, the set of participants for a particular group is fixed and defined at group creation.

    \item \textbf{Server Timestamps:} The server is trusted to add the current timestamp to every message.

\end{itemize}%

Under the assumption that the server faithfully delivers encrypted messages to all group participants using the Sender Key protocol via server fan-out, we can implicitly ensure that every group participant receives the same (signed) ciphertexts. 
There remain two cases that can still be problematic:

\paragraph{Different Key with Valid Decryption.}
Since each participant distributes their symmetric sender key to every other participant individually, a malicious sender can provide different recipients with different sender keys. 
In principle, a malicious sender could attempt to find a ciphertext and key combination that meaningfully decrypts into two different plaintexts.
This attack was discussed in~\cite{schliep_end--end_2019} and its success probability can be considered negligible with adequate parameter settings and domain separation.

\paragraph{Retransmission Due to Decryption Failure.}
Because group membership changes and device state resets can occur at any time, a participant may be unable to decrypt a received message using its current sender key for a given sender. %
This is an inherent property of the protocol and the asynchronous communication model.
A malicious participant can abuse this by distributing different sender keys to trigger retry requests with the fallback to pairwise 1:1 communication, providing the original sender with the possibility to equivocate.

The first, albeit insufficient, countermeasure that suggests itself is to require that any retransmitted message also be redistributed to the entire group via server fan-out; we explain below why this approach is inadequate.
Participants who already received and decrypted the message would then need to verify that the retransmitted ciphertext decrypts successfully and that its plaintext matches their local copy. 
If decryption fails again for a participant, that participant would need to issue a retry request for the retransmission itself.
In effect, recipients would request retries for messages that already appear in their transcript solely to validate that the retransmitted content matches, which introduces a complex error resolution loop with no clear abort cutoff point. 

To avoid this complexity, we propose treating \emph{any} decryption error or retransmission as a potential equivocation signal.
Rather than attempting to fix retransmissions within the current protocol, we suggest surfacing decryption errors and retransmissions in the UI, together with an explicit warning that the retransmitted content might differ across participants, and providing decryption error and retransmission statistics on a per-participant basis.
Such statistics help distinguish false positives from attacks and support identifying the attacker. %

\subsubsection{Guarantees and Limitations}
If no decryption failures occur, all participants are guaranteed to eventually see the same messages (STC). In the presence of decryption failures and associated warnings, no guarantees can be provided for the content of the affected messages.
However, highlighting such messages in the UI as potentially different across participants 
can already act as a sufficient deterrent against targeted equivocation.

Currently, the ordering of messages works by appending messages as received, or based on when a message was sent from the local device. If messages would be ordered by server timestamp, this could lead to cases where messages ``hop around'' in the chat window: For example, a message created locally would require a roundtrip to the server to get a server timestamp according to which it can be ordered in the UI. The alternative would be to only reorder messages totally (TTC) in retrospect upon explicit user intent.

While a malicious participant could deliberately trigger warning messages, our countermeasure still flags the anomaly and exposes the involved sender and recipient, thus working as intended. This improves upon the status quo, where attackers can operate completely undetected. A more pressing concern is false positives caused by legitimate retransmissions due to technical failures. A short self-study did not reveal any decryption failures during normal operation, but accurately quantifying their real-world likelihood would require a large-scale experiment and a complementary user study that are difficult to deploy without platform-operator support. Thus, this is left for future work. More fundamentally, a general, protocol-agnostic fix is a hard problem, as prior approaches rely on full protocol redesigns and often neglect performance. Rather than attempting such a redesign in this paper, which would exceed typical conference length constraints, we deliberately present the least intrusive mitigation that integrates into deployed systems, fully aware of its imperfections.

\subsubsection{Implementation Considerations}
We highlight the cornerstones of the proposed changes for WhatsApp and Signal to account for the different protocol architectures.
The two designs differ in the server's view of group membership.
In WhatsApp, the server is aware of the group participants, whereas Signal supports private groups where the server does not store the group composition in plaintext.
To account for our previous \emph{Enforced Group Broadcast} assumption, the following outlines how this may be achievable with the respective messenger services.

\paragraph{WhatsApp.}
In the case of WhatsApp, the server can check and ensure delivery to the entire group by directly invoking the appropriate group endpoint and relying on the group state information already available on the WhatsApp server.

\paragraph{Signal.}
Signal aims to minimize the plaintext state and metadata available on its servers, for example by supporting private groups~\cite{marlinspike_private_2014,chase_signal_2020} and sealed sender~\cite{signal_technology_foundation__signalapp_libsignal-protocol-java_2022,martiny_improving_2021}.
While the chat server's validation process for multi-recipient group sends checks the supplied group send token for valid endorsements against the list of recipients~\cite{signalserver_groupsendtoken_2026}, the current zero-knowledge architecture does not seem to allow for direct associations with group member state~\cite{libsignal_groupsendendorsements_2025}.
As a consequence, the chat server cannot directly verify that a multi-recipient message targets all group participants, which necessitates shifting delivery checks to the clients.

In this setting, the server needs to attach some form of a participant list\footnote{For example, a hash of the participant list.} to each multi-recipient message\footnote{Which must not be a story, i.e., have the \emph{isStory} flag set.}.
In return, this approach enables validation of group messages against the authoritative group state maintained via \emph{zkgroups}.
Concretely, clients could validate the server-provided recipient list against the group state referenced in the message plaintext through its \emph{group revision number}.
This requires that the correct revision number is added by the sending client, and that the receiving client is able to reconstruct the according participant list. 
Since group membership changes could occur at any time, these client-side checks may conflict with concurrent changes to group state. In the simplest case, group membership is static and defined at creation\footnote{Alternatively, a sending client providing an incorrect revision number could be detected by relying on server-provided timestamps for messages and recorded group state changes. The sender would then verify the revision using the server timestamp.}.
If the reconstructed participant list does not match the information provided by the server, receiving clients should drop the respective message.

\section{Related Work and TC History}
\label{sec:related_work}

One of the earliest works on transcript agreement in group messaging is by Goldberg et al.~\cite{goldberg_multi-party_2009}, who proposed Multi-party Off-the-Record Messaging (mpOTR).
They described a property closely related to \ac{TC}, which they termed \emph{consensus}.
Their approach relies on retrospective consistency checks after session termination, allowing participants to detect violations but offers no remediation during an ongoing conversation.
Marlinspike later adopted this idea when informally describing a desired consistency property for TextSecure~\cite{marlinspike_private_2014}, which later evolved into Signal.
As we demonstrate, current Signal group chats do not provide this property.

Unger et al.~\cite{unger_sok_2015} systematized transcript-related guarantees and decomposed \ac{TC} into three conceptually related desiderata, namely \emph{speaker consistency}, \emph{causality preserving}, and a \emph{global transcript}.
They emphasized that a global transcript implies speaker consistency and further observed that pairwise OTR connections in groups inherently violate both speaker consistency and causality.

Subsequent work shifted focus toward attacks by a malicious messaging server, while explicitly or implicitly assuming that all group participants behave honestly.
Schliep et al.~\cite{schliep_is_2017} analyzed attacks caused by a malicious Signal server, with a focus on message reordering and message dropping.
They did not consider inconsistencies caused by malicious group members.
A follow-up work in 2019~\cite{schliep_end--end_2019} proposed conversation integrity guarantees by introducing order-enforcing service providers that act similarly to trusted timestamping services.
From a practical perspective, these results do not directly transfer to modern Signal group chats, since the implementation at the time differed substantially from today's design and did not yet support the Sender Key protocol~\cite{whatsapp_whatsapp_2023} or private groups~\cite{jimio_technology_2019}.

Research on the Sender Key protocol itself focused on formal modeling and security analysis~\cite{balbas_analysis_2023} and on protocol improvements~\cite{balbas_whatsupp_2023,chase_signal_2020}.
These works do not address \ac{TC} violations caused by equivocation or selective message delivery by malicious participants.

R{\"o}sler et al.~\cite{rosler_more_2018} formalized security goals for group messaging and introduced \emph{traceable delivery}, which allows detecting whether a sent message was received.
Their threat model assumes group members to be honest and to follow the protocol.
At the time of their study, Signal did not provide this property, not even for 1:1 chats.
In contrast, we explicitly consider malicious group participants and focus on inconsistencies that arise even when the server behaves correctly.
Several works studied properties closely related to \ac{TC}, such as causality and history integrity.
Eugster et al.~\cite{eugster_cryptographic_2018} explored goals for secure group communication while considering causality, but they neither aimed to tolerate Byzantine participants nor required total order delivery.
Barooti et al.~\cite{barooti_active_2023} introduced \emph{history integrity} and studied active attacks in the presence of a compromised device state.
Their analysis focused on 1:1 conversations and did not consider group messaging.
Transcript franking~\cite{chen_integrating_2024,namavari_transcript_2025} represents another related line of work.
It aims to enable selective disclosure of conversation transcripts to moderators or authorities in cases of abuse.
Correct transcript franking requires preserving message causality.
Chen et al.~\cite{chen_integrating_2024} showed that existing instant messengers fail to preserve causality even in 1:1 conversations, and that TLS~1.3 itself does not guarantee this property.
They further identified causality-preserving group messaging as an open problem.

MLS~\cite{barnes_messaging_2023,beurdouche_messaging_2025,robert_messaging_2025} defines a notion termed \emph{transcript consistency}, but this property applies exclusively to the handshake protocol and thus to group state rather than application messages~\cite{robert_messaging_2025-1}.
MLS aims to ensure that all honest members share a consistent view of group state, including group membership, proposal processing, and key schedule evolution.
However, MLS is a group key establishment and state agreement protocol, not a messaging protocol.
It deliberately leaves \ac{TC} for application messages to the application layer.

Recent work has shown that legitimate users can exploit protocol mechanisms in deployed E2EE messengers to undermine security and privacy guarantees\cite{gegenhuber_2024_carelesswhisper,gegenhuber_2025_prekeypogo,garske_2026_blue}.
Complementing these lines of work, we study \ac{TC} violations caused by malicious group participants in deployed \ac{E2EE} messaging applications, demonstrate practical attacks, and discuss mitigations that integrate with the Sender Key protocol.

\section{Discussion}
While confidentiality and authenticity against external adversaries are relatively well-understood and carefully engineered, the integrity of the shared group transcript in the presence of malicious participants has largely remained unaddressed.

This gap is particularly visible from the user's perspective.
Group chats (and especially interactive features such as polls) create the expectation that all participants observe the same conversation and derive the same outcomes.
In practice, none of the analyzed systems enforce such guarantees.
As a result, \ac{TC} failures violate basic user expectations about how group communication functions.

A key reason for this mismatch is the prevailing threat model in protocol design, which does not account for participants targeting the consistency of the conversation. %
Over the last decade, encrypted messaging applications have shifted from services for communication among trusted parties (e.g., friends and family) to platforms that support large, open groups where participants join via links or QR codes, representing much weaker trust relationships. Additionally, they are used as channels for coordination and decision-making by organizations, officials, and other high-value groups in adversarial settings (e.g., the current geopolitical situation). In this context, the threat does not necessarily stem from malicious intent on the part of an account holder but may instead arise from a threat actor who has compromised a user’s device participating in the respective group.

Beyond impact for normal messages, additional message types (e.g., location sharing, polls) create even more attack vectors and further amplify the problem.
We observed numerous behaviors (e.g., naive implementations of polls, platform-dependent message handling) that reduce attacker effort and increase the practical impact of transcript inconsistencies.
The prevalence of such issues stands in contrast to the expectation of highly hardened messengers deployed at a global scale.

A natural direction for future work is to better understand how transcript inconsistencies (or future UI warnings) manifest at the user level and whether they are noticed in practice.
Extending our analysis to additional E2EE messaging platforms and emerging group messaging designs (e.g., MLS) could help assess whether \ac{TC} can be readily improved in newer systems or whether similar design trade-offs persist.
Finally, we argue that \ac{TC} should be addressed at the design level with the same rigor as other well-established security and privacy properties of modern E2EE instant messengers.

\section{Conclusion}
In this paper, we answered \emph{What is the current state of \ac{TC} in \ac{E2EE} messaging?} by showing that all analyzed \ac{E2EE} messengers, namely WhatsApp, Signal, iMessage, and Threema, failed to provide \ac{TC}.
As a result, malicious group participants could selectively omit, reorder, or tailor messages to present different content to different participants.
These attacks directly undermine the integrity of the shared group transcript.

Our analysis showed that these weaknesses arose from fundamental design choices rather than isolated implementation bugs.
Despite long-standing awareness of transcript consistency issues in academia and expert communities, current systems offer neither preventive mechanisms nor effective means to detect them (neither on the server nor on the receiving client).
The risk posed by transcript inconsistency is further amplified by growing group sizes, newly introduced features such as polls, and the sensitive contexts in which these messengers are used in current times.
To improve the existing situation, we proposed practical protocol changes that require minimal adaptations to the Sender Key protocol, which Signal and WhatsApp use, to detect inconsistencies.

\section*{Ethical Considerations}

Overall, we conducted all experiments exclusively using our own accounts to avoid any harm to real users. 
We further identify two stakeholder groups affected by our findings: (1) instant messaging users, and (2) platform operators. In the following, we discuss the potential impact of our findings on each stakeholder group and describe the mitigation measures we apply to reduce harm.

\paragraph{Instant Messaging Users.}
Publishing our research could facilitate the exploitation of \ac{TC} inconsistencies. Attackers could leverage our insights for sophisticated social engineering attacks against instant messaging users.

However, withholding our findings would amount to security by obscurity, which practical experience has repeatedly shown to be ineffective as a security strategy. %
We include our \ac{PoC} code for all messenger applications in the paper submission and publish it to support reproducibility and future work after a 90 days grace period.

The vulnerabilities affect all analyzed \ac{E2EE} messaging apps, which makes raising public awareness essential. Increased awareness can help users to better understand the threat model and security guarantees of group chats, or even recognize and detect such attacks in certain cases. In addition, disseminating our findings within the research community fosters the development and evaluation of countermeasures.

Additionally, we aimed to minimize potential harm to users and platforms by responsibly disclosing our findings to the affected platform operators before publishing. 

\paragraph{Platform Operators.}

Highlighting existing issues is essential to enable long-term improvements. 
Our proposed countermeasures require only minimal design changes to the existing Sender Key protocol, which WhatsApp and Signal currently use. Therefore, we consider our countermeasures a practical route to improve upon the current situation with relatively low overhead. 
In addition, it is particularly important to incorporate our findings into future protocol designs, as mitigating these risks without understanding them is challenging. 

As mentioned earlier, we responsibly disclosed all findings to the platform operators of the analyzed messaging apps, which gives them the opportunity to address the issues before public disclosure.

\section*{Responsible Disclosure}
On February 6th, 2026 we responsibly disclosed the identified issues to the affected platform operators by sharing a preprint of this paper together with individual issue reports.
For WhatsApp, Apple, and Threema, we used the respective bug bounty or vulnerability disclosure platforms.
Signal was contacted via its (publicly recommended) security contact email address.

At the time of writing (2026-07-29), all messengers (WhatsApp, Apple, Threema, and Signal) acknowledged receipt of the reports, and provided feedback on the identified issues.
Meta awarded a USD 1,000 bug bounty and indicated plans to address implementation-specific issues (e.g., adding security hashes protecting the poll question), but no protocol-level transcript consistency changes were communicated.
Similarly, Apple expressed appreciation for the disclosure, but indicated that it does not currently plan changes related to transcript consistency.
Threema plans to address implementation-specific issues (i.e., inflated votes by a single participant), while they consider fabricated results of a poll creator less relevant to their current threat model and also do not plan to implement protocol-level changes regarding transcript consistency in the near future.
Signal highlighted that transcript consistency does not currently constitute an explicit security goal. However, Signal also noted that its ongoing redesign of group chats may bring transcript consistency closer to its scope of future security goals.

\paragraph{Response Time.}
Apple responded approximately one month after the disclosure, while WhatsApp and Threema each responded after roughly three months, and Signal took about five months to respond.

\paragraph{Paper and Artifact Release.}
We made sure to give all platform operators sufficient time ($>$ 90 days) to process our reports, communicate their mitigation plans and request an embargo.
To support comprehensive evaluation and reproducibility of our work and to increase awareness within the community, we publicly release the artifacts at \url{https://github.com/sbaresearch/transcript-consistency}.

\section*{Open Science}
Since more than 90 days have passed and vendors do not consider TC as part of their threat model, the artifacts are made available to foster reproducibility and future work: on GitHub at \url{https://github.com/sbaresearch/transcript-consistency} and on Zenodo for long-term archiving at \url{https://doi.org/10.5281/zenodo.20323807}.

The provided repository contains the following:

\begin{itemize}
    \item \texttt{code/}: This directory contains the code of our custom clients for all
    targeted messengers.
    \item \texttt{code/imessage/}: This directory contains a modified version of rustpush~\cite{rustpush} that includes a command-line client to send inconsistent messages into group chats and to manipulate polls and votes.
    \item \texttt{code/signal/poc-skdm/}: This directory contains a modified Signal Desktop~\cite{signal_desktop}
    client demonstrating the Sender Key with E2EE fallback case for a specific recipient
    selected in the GUI.
    \item \texttt{code/signal/poc-groups/}: This directory contains a modified Signal Desktop~\cite{signal_desktop}
    client supporting server and client fan-out-only modes and providing a more advanced GUI for
    exclusions and overrides for all considered message types (e.g., polls, regular messages).
    \item \texttt{code/threema/threema-android/}: This directory contains a modified Threema
    Android~\cite{threema_android} client with support for commands in relevant input fields (for regular messages and polls) to control excluded recipients and overrides, as well as poll voting and closing.
    \item \texttt{code/whatsapp/poc-client/}: This directory contains our modified WhatsApp client that allows sending inconsistent  messages while supporting server and client fan-out-only modes, as well as Sender Key with E2EE fallback, which builds on whatsmeow~\cite{whatsmeow}.
    \item \texttt{code/whatsapp/quote-client/}: This directory contains our modified WhatsApp client to manipulate quotes, building on whatsmeow~\cite{whatsmeow}.
    \item \texttt{videos/}: This directory contains \ac{PoC} videos showing the
    behavior of official clients in a group with a malicious client.
\end{itemize}

Throughout the repository, \texttt{README.md} files provide the information
required to reproduce the results.

\paragraph{Environment.}

For the Signal Desktop client, in particular, we
provide containers for improved reproducibility since there are many dependencies
and the build steps would otherwise be prone to failure. The entire GUI can be run inside
a container.

\section*{Acknowledgments}
The financial support by the Austrian Federal Ministry of Economy, Energy and Tourism, the National Foundation for Research, Technology and Development, the Christian Doppler Research Association, the FFG Bridge project 46322124 SecKey, the FFG KIRAS/K-PASS project 59103683 TelCrit, and the University of Vienna, Faculty of Computer Science, Security \& Privacy Group, SBA Research (SBA-K1 NGC) a COMET Center within the COMET – Competence Centers for Excellent Technologies Program funded by BMIMI, BMWET, and the federal state of Vienna, is gratefully acknowledged.
We would also like to thank our anonymous reviewers for their valuable feedback and suggestions.

\bibliographystyle{plainurl}
\bibliography{bib/bibliography,bib/messaging}

@misc{gegenhuber_2024_carelesswhisper,
	title        = {{Careless Whisper: Exploiting Silent Delivery Receipts to Monitor Users on Mobile Instant Messengers}},
	author       = {Gabriel K. Gegenhuber and Maximilian Günther and Markus Maier and Aljosha Judmayer and Florian Holzbauer and Philipp {\'E}. Frenzel and Johanna Ullrich},
	year         = {2024},
	url          = {https://arxiv.org/abs/2411.11194},
	eprint       = {2411.11194},
	archiveprefix = {arXiv},
	primaryclass = {cs.CR}
}

@techreport{threema_threema_2025,
	title = {{Cryptography Whitepaper}},
	url = {https://threema.com/press-files/2_documentation/cryptography_whitepaper.pdf},
	urldate = {2026-01-30},
	author = {{Threema}},
	month = mar,
	year = {2025},
    note = {Accessed: 2026-01-30},
}

@misc{mau_signal,
  title = {{Signal Statistics}},
  url   = {https://www.businessofapps.com/data/signal-statistics/},
  note  = {Accessed: 2026-01-23}
}

@misc{mau_imessage,
  title = {{iMessage Statistics}},
  url   = {https://usesignhouse.com/blog/imessage-stats/},
  note  = {Accessed: 2026-01-23}
}

@misc{mau_threema,
  title = {{Threema Press Information}},
  url   = {https://threema.com/press-files/1_press_info/press_threema_portrait_en.pdf},
  note  = {Accessed: 2026-01-23}
}

@inproceedings{gegenhuber_2025_prekeypogo,
    author    = {Gegenhuber, Gabriel K. and Frenzel, Philipp {\'E}. and G{\"u}nther, Maximilian and Jud
ayer, Aljosha},
    title     = {{Prekey Pogo: Investigating Security and Privacy Issues in WhatsApp's Handshake Mechan
sm}},
    booktitle = {19th USENIX WOOT Conference on Offensive Technologies (WOOT)},
    year      = {2025}
}

@inproceedings{gegenhuber_2025_heythere,
    author    = {Gegenhuber, Gabriel K. and Frenzel, Philipp {\'E}. and G{\"u}nther, Maximilian and Ullrich, Johanna and Judmayer, Aljosha},
    title     = {{Hey there! You are using WhatsApp: Enumerating Three Billion Accounts for Security and Privacy}},
    booktitle = {33rd Annual Network and Distributed System Security Symposium (NDSS)},
    year      = {2026}
}

@inproceedings{garske_2026_blue,
  title={{Blue Bubbles, Red Flags: Investigating Privacy Leakage in Apple iMessage}}, 
  author={Viktor E. Garske and Swantje Lange and Gabriel K. Gegenhuber and David Schmidt and Andreas Noack and Jiska Classen},
  booktitle={Proceedings of the 2026 ACM SIGSAC Conference on Computer and Communications Security},
  year={2026}
}

@misc{signal_support_group_chats,
  author       = {{Signal Support}},
  title        = {{Group chats}},
  url          = {https://support.signal.org/hc/en-us/articles/360007319331-Group-chats},
  note         = {Accessed: 2026-01-27},
  organization = {Signal Foundation}
}

@misc{threema_android,
	title        = {{GitHub -- Threema for Android}},
	url          = {https://github.com/threema-ch/threema-android},
	note         = {Accessed: 2026-01-27},
}

@misc{whatsmeow,
	title        = {{GitHub -- whatsmeow}},
	url          = {https://github.com/tulir/whatsmeow},
	note         = {Accessed: 2026-01-27},
}

@misc{signal_desktop,
	title        = {{GitHub -- Signal Desktop}},
	url          = {https://github.com/signalapp/Signal-Desktop},
	note         = {Accessed: 2026-01-27},
}

@misc{rustpush,
	title        = {{GitHub -- Rustpush}},
	url          = {https://github.com/OpenBubbles/rustpush},
	note         = {Accessed: 2026-01-27},
}

@misc{threema_polls,
	title        = {{Threema Poll Feature}},
    author       = {{Threema}},
	url          = {https://threema.com/en/blog/threema-poll-feature},
    date         = {2015-01-12},
	note         = {Accessed: 2026-01-27},
}

@misc{whatsapp_2022_polls,
	title        = {{WhatsApp Revamps Group Chats with New Communities Feature}},
    author = {Thomas Germain},
	url          = {https://gizmodo.com/whatsapp-communities-new-group-chat-feature-1849737258},
    date = {2022-11-03},
	note         = {Accessed: 2026-01-27},
}

@misc{whatsapp_2023_polls,
	title        = {{New Updates to Polls and Sharing With Captions on WhatsApp}},
    author      = {{Meta}},
	url          = {https://about.fb.com/news/2023/05/whatsapp-polls-updates-sharing-with-captions},
    date        = {2023-05-04},
	note         = {Accessed: 2026-01-27},
}

@misc{imessage_polls,
	title        = {{Apple is bringing polls to Messages in iOS 26}},
    author       = {Aisha Malik},
	url          = {https://techcrunch.com/2025/06/09/apple-is-bringing-polls-to-imessage-in-ios-26},
    date         = {2025-06-09},
	note         = {Accessed: 2026-01-27},
}

@misc{signal_polls,
	title        = {{Signal Polls: Yes, no, maybe (yes!)}},
    author       = {Nina Berman},
	url          = {https://signal.org/blog/polls},
    date         = {2025-11-19},
	note         = {Accessed: 2026-01-27},
}

@misc{whatsapp_group_size,
	title        = {{WhatsApp -- How to join a group in a community}},
    author      = {{Meta}},
	url          = {https://faq.whatsapp.com/967457667545238/},
	note         = {Accessed: 2026-01-27},
}

@misc{iMessage_group_size,
	title        = {{Apple Community -- Sending group messages to 50+ recipients}},
	url          = {https://discussions.apple.com/thread/255951019},
	note         = {Accessed: 2026-01-27},
}

@misc{threema_group_size,
	title        = {{Threema for iOS: Larger Groups and More}},
    author      = {{Threema}},
	url          = {https://threema.com/en/blog/threema-464-for-ios},
	note         = {Accessed: 2026-01-27},
}

@misc{threema,
	title        = {{Secure Communication For Individuals and Companies}},
    author      = {{Threema}},
	url          = {https://threema.com/en},
	note         = {Accessed: 2026-01-27},
}

@misc{whatsapp,
	title        = {{Secure and Reliable Free Private Messaging and Calling}},
    author      = {{WhatsApp}},
	url          = {https://www.whatsapp.com/},
	note         = {Accessed: 2026-01-27},
}

@misc{iMessage,
	title        = {{Messages for iPhone, iPad, Apple Watch, and Mac}},
    author      = {{Apple}},
	url          = {https://support.apple.com/messages},
	note         = {Accessed: 2026-01-27},
}

@misc{signal,
	title        = {{Signal >> Home}},
    author      = {{Signal}},
	url          = {https://signal.org/},
	note         = {Accessed: 2026-01-27},
}

@misc{wa_win_webview,
  title = {{Meta just killed native WhatsApp on Windows 11, now it opens WebView}},
  url   = {https://www.windowslatest.com/2025/11/12/meta-just-killed-native-whatsapp-on-windows-11-now-it-opens-webview-uses-1gb-ram-all-the-time/},
  note  = {Accessed 2026-01-23}
}

@misc{frida,
	title        = {{Frida --  A world-class dynamic instrumentation toolkit}},
	author       = {Ravnås, Ole André V.},
	url          = {https://frida.re/}
}

@misc{signalandroid_remoteconfig_2026,
  author       = {{signalapp}},
  title        = {{RemoteConfig.kt} at commit \texttt{1ddde6a} in \textit{Signal-Android}},
  year         = {2026},
  note         = {Accessed: 2026-02-05},
  organization = {GitHub},
  url          = {https://github.com/signalapp/Signal-Android/blob/1ddde6ab92f39dff0987f60a08460412a0879a55/app/src/main/java/org/thoughtcrime/securesms/util/RemoteConfig.kt#L547-L563}
}

@misc{signalandroid_sender_key,
  author       = {{signalapp}},
  title        = {{Signal-Android: Commit 0459d118a397e5405c9742e3f6238189e1fe0c9d}},
  date         = {2021-08-25},
  note         = {Accessed: 2026-02-05},
  organization = {GitHub},
  url          = {https://github.com/signalapp/Signal-Android/commit/0459d118a397e5405c9742e3f6238189e1fe0c9d}
}

@misc{signalserver_acceptedtypes_2025,
  author       = {{signalapp}},
  title        = {{IncomingMessage.java} at commit \texttt{dc3920a} in \textit{Signal-Server}},
  year         = {2025},
  note         = {Accessed: 2026-02-04},
  organization = {GitHub},
  url          = {https://github.com/signalapp/Signal-Server/blob/dc3920a99cb25cae66ca7439004545b69ae55ca4/service/src/main/java/org/whispersystems/textsecuregcm/entities/IncomingMessage.java#L79-L89}
}

@misc{libsignal_groupsendendorsements_2025,
  author       = {{signalapp}},
  title        = {{GroupSendEndorsement.ts} at commit \texttt{85686ca} in \textit{libsignal}},
  year         = {2025},
  note         = {Accessed: 2026-02-05},
  organization = {GitHub},
  url          = {https://github.com/signalapp/libsignal/blob/85686caa01465eacba6fddcdc19a22d2d62d8c7f/node/ts/zkgroup/groupsend/GroupSendEndorsement.ts}
}

@misc{signalserver_groupsendtoken_2026,
  author       = {{signalapp}},
  title        = {{MessageController.java} at commit \texttt{ad21f00} in \textit{Signal-Server}},
  year         = {2026},
  note         = {Accessed: 2026-02-05},
  organization = {GitHub},
  url          = {https://github.com/signalapp/Signal-Server/blob/ad21f002ab837f931c28a5ea020d82eb0b1f43aa/service/src/main/java/org/whispersystems/textsecuregcm/controllers/MessageController.java#L589}
}

@misc{threema_ios_info_plist_2025,
  author       = {{threema-ch}},
  title        = {{Threema-Info.plist} at commit \texttt{7b1636e} in \textit{threema-ios}},
  year         = {2025},
  note         = {Accessed: 2026-01-29; iOS app source file from the Threema open-source repository.},
  organization = {GitHub},
  url          = {https://github.com/threema-ch/threema-ios/blob/7b1636e9a1e765f6aa1db7a71420c4d3874b0ca5/Threema/SupportingFiles/Threema/Threema-Info.plist#L118-L119}
}

@misc{signal_video_14,
    author = {Gabriel K. Gegenhuber and Moritz Grefner and Maximilian Günther and Matthäus Wininger and David Schmidt and Aljosha Judmayer},
    title = {{Malicious Signal Message Referencing Future Message}},
    url = {https://github.com/sbaresearch/transcript-consistency/blob/main/videos/signal/14/fake-quote-2-wrong-causality.mkv}
}

@misc{signal_video_15,
    author = {Gabriel K. Gegenhuber and Moritz Grefner and Maximilian Günther and Matthäus Wininger and David Schmidt and Aljosha Judmayer},
    title = {{Duplicate Message ID in Signal}},
    url = {https://github.com/sbaresearch/transcript-consistency/blob/main/videos/signal/15/desktop-duplicate-handling-1.mkv}
}

@misc{signal_video_16,
    author = {Gabriel K. Gegenhuber and Moritz Grefner and Maximilian Günther and Matthäus Wininger and David Schmidt and Aljosha Judmayer},
    title = {{Duplicate Message Edit ID in Signal}},
    url = {https://github.com/sbaresearch/transcript-consistency/blob/main/videos/signal/16/desktop-duplicate-handling-2.mkv}
}

@misc{marlinspike_private_2014,
	title = {Private {Group} {Messaging}},
	url = {https://signal.org/blog/private-groups/},
	urldate = {2024-08-06},
	author = {Marlinspike, Moxie},
	month = may,
	year = {2014},
}

@inproceedings{rosler_more_2018,
	title = {More is {Less}: {On} the {End}-to-{End} {Security} of {Group} {Chats} in {Signal}, {WhatsApp}, and {Threema}},
	url = {https://doi.org/10.1109/EuroSP.2018.00036},
	doi = {10.1109/EUROSP.2018.00036},
	booktitle = {2018 {IEEE} {European} {Symposium} on {Security} and {Privacy}, {EuroS}\&{P} 2018, {London}, {United} {Kingdom}, {April} 24-26, 2018},
	publisher = {IEEE},
	author = {R{\"o}sler, Paul and Mainka, Christian and Schwenk, J{\"o}rg},
	year = {2018},
	pages = {415--429},
}

@techreport{whatsapp_whatsapp_2023,
	title = {{WhatsApp} {Encryption} {Overview}: {Technical} white paper},
	url = {https://www.whatsapp.com/security/WhatsApp-Security-Whitepaper.pdf},
	urldate = {2024-08-06},
	author = {Whatsapp},
	month = sep,
	year = {2023},
}

@inproceedings{chase_signal_2020,
	title = {The {Signal} {Private} {Group} {System} and {Anonymous} {Credentials} {Supporting} {Efficient} {Verifiable} {Encryption}},
	url = {https://eprint.iacr.org/2019/1416.pdf},
	doi = {10.1145/3372297.3417887},
	booktitle = {{CCS} '20: 2020 {ACM} {SIGSAC} {Conference} on {Computer} and {Communications} {Security}, {Virtual} {Event}, {USA}, {November} 9-13, 2020},
	publisher = {ACM},
	author = {Chase, Melissa and Perrin, Trevor and Zaverucha, Greg},
	editor = {Ligatti, Jay and Ou, Xinming and Katz, Jonathan and Vigna, Giovanni},
	year = {2020},
	pages = {1445--1459},
}

@inproceedings{schliep_is_2017,
	title = {Is {Bob} {Sending} {Mixed} {Signals}?},
	url = {https://www-users.cse.umn.edu/~hoppernj/mixed_signals_wpes17.pdf},
	doi = {10.1145/3139550.3139568},
	booktitle = {Proceedings of the 2017 on {Workshop} on {Privacy} in the {Electronic} {Society}, {Dallas}, {TX}, {USA}, {October} 30 - {November} 3, 2017},
	publisher = {ACM},
	author = {Schliep, Michael and Kariniemi, Ian and Hopper, Nicholas},
	editor = {Thuraisingham, Bhavani and Lee, Adam J.},
	year = {2017},
	pages = {31--40},
}

@misc{jimio_technology_2019,
	title = {Technology {Preview}: {Signal} {Private} {Group} {System}},
	url = {https://signal.org/blog/signal-private-group-system/},
	author = {jimio},
	month = dec,
	year = {2019},
}

@misc{signal_messenger_new_2020,
	title = {New {Features} {Coming} to {Signal} {Groups}},
	url = {https://signal.org/blog/new-groups/},
	author = {{Signal Messenger}},
	month = oct,
	year = {2020},
}

@article{balbas_analysis_2023,
	title = {Analysis and {Improvements} of the {Sender} {Keys} {Protocol} for {Group} {Messaging}},
	volume = {abs/2301.07045},
	url = {https://doi.org/10.48550/arXiv.2301.07045},
	doi = {10.48550/ARXIV.2301.07045},
	journal = {CoRR},
	author = {Balb{\'a}s, David and Collins, Daniel and Gajland, Phillip},
	year = {2023},
	note = {arXiv: 2301.07045},
}

@inproceedings{balbas_whatsupp_2023,
	series = {Lecture {Notes} in {Computer} {Science}},
	title = {{WhatsUpp} with {Sender} {Keys}? {Analysis}, {Improvements} and {Security} {Proofs}},
	volume = {14442},
	url = {https://doi.org/10.1007/978-981-99-8733-7\_10},
	doi = {10.1007/978-981-99-8733-7_10},
	booktitle = {Advances in {Cryptology} - {ASIACRYPT} 2023 - 29th {International} {Conference} on the {Theory} and {Application} of {Cryptology} and {Information} {Security}, {Guangzhou}, {China}, {December} 4-8, 2023, {Proceedings}, {Part} {V}},
	publisher = {Springer},
	author = {Balb{\'a}s, David and Collins, Daniel and Gajland, Phillip},
	editor = {Guo, Jian and Steinfeld, Ron},
	year = {2023},
	pages = {307--341},
}

@inproceedings{kleppmann_secure_2018,
	series = {Lecture {Notes} in {Computer} {Science}},
	title = {From {Secure} {Messaging} to {Secure} {Collaboration}},
	volume = {11286},
	url = {https://martin.kleppmann.com/papers/secure-collaboration-spw18.pdf},
	doi = {10.1007/978-3-030-03251-7_21},
	booktitle = {Security {Protocols} {XXVI} - 26th {International} {Workshop}, {Cambridge}, {UK}, {March} 19-21, 2018, {Revised} {Selected} {Papers}},
	publisher = {Springer},
	author = {Kleppmann, Martin and Kollmann, Stephan A. and Vasile, Diana A. and Beresford, Alastair R.},
	editor = {Maty{\'a}s, Vashek and Svenda, Petr and Stajano, Frank and Christianson, Bruce and Anderson, Jonathan},
	year = {2018},
	pages = {179--185},
}

@inproceedings{namavari_transcript_2025,
	series = {Lecture {Notes} in {Computer} {Science}},
	title = {Transcript {Franking} for {Encrypted} {Messaging}},
	volume = {16246},
	url = {https://doi.org/10.1007/978-981-95-5096-8\_1},
	doi = {10.1007/978-981-95-5096-8_1},
	booktitle = {Advances in {Cryptology} - {ASIACRYPT} 2025 - 31st {International} {Conference} on the {Theory} and {Application} of {Cryptology} and {Information} {Security}, {Melbourne}, {VIC}, {Australia}, {December} 8-12, 2025, {Proceedings}, {Part} {II}},
	publisher = {Springer},
	author = {Namavari, Armin and Ristenpart, Thomas},
	editor = {Hanaoka, Goichiro and Yang, Bo-Yin},
	year = {2025},
	pages = {3--33},
}

@inproceedings{cohn-gordon_ends--ends_2018,
	title = {On {Ends}-to-{Ends} {Encryption}: {Asynchronous} {Group} {Messaging} with {Strong} {Security} {Guarantees}},
	url = {https://doi.org/10.1145/3243734.3243747},
	doi = {10.1145/3243734.3243747},
	booktitle = {Proceedings of the 2018 {ACM} {SIGSAC} {Conference} on {Computer} and {Communications} {Security}, {CCS} 2018, {Toronto}, {ON}, {Canada}, {October} 15-19, 2018},
	publisher = {ACM},
	author = {Cohn-Gordon, Katriel and Cremers, Cas and Garratt, Luke and Millican, Jon and Milner, Kevin},
	editor = {Lie, David and Mannan, Mohammad and Backes, Michael and Wang, XiaoFeng},
	year = {2018},
	pages = {1802--1819},
}

@inproceedings{unger_sok_2015,
	title = {{SoK}: {Secure} {Messaging}},
	url = {https://doi.org/10.1109/SP.2015.22},
	doi = {10.1109/SP.2015.22},
	booktitle = {2015 {IEEE} {Symposium} on {Security} and {Privacy}, {SP} 2015, {San} {Jose}, {CA}, {USA}, {May} 17-21, 2015},
	publisher = {IEEE Computer Society},
	author = {Unger, Nik and Dechand, Sergej and Bonneau, Joseph and Fahl, Sascha and Perl, Henning and Goldberg, Ian and Smith, Matthew},
	year = {2015},
	pages = {232--249},
}

@inproceedings{chen_integrating_2024,
	series = {Lecture {Notes} in {Computer} {Science}},
	title = {Integrating {Causality} in {Messaging} {Channels}},
	volume = {14653},
	url = {https://eprint.iacr.org/2024/362.pdf},
	doi = {10.1007/978-3-031-58734-4_9},
	booktitle = {Advances in {Cryptology} - {EUROCRYPT} 2024 - 43rd {Annual} {International} {Conference} on the {Theory} and {Applications} of {Cryptographic} {Techniques}, {Zurich}, {Switzerland}, {May} 26-30, 2024, {Proceedings}, {Part} {III}},
	publisher = {Springer},
	author = {Chen, Shan and Fischlin, Marc},
	editor = {Joye, Marc and Leander, Gregor},
	year = {2024},
	pages = {251--282},
}

@misc{signal_technology_foundation__signalapp_libsignal-protocol-java_2022,
	title = {libsignal-protocol-java: {A} {Java} implementation of the {Signal} {Protocol}},
	url = {https://github.com/signalapp/libsignal-protocol-java},
	author = {{Signal Technology Foundation / signalapp}},
	year = {2022},
}

@inproceedings{barooti_active_2023,
	series = {Lecture {Notes} in {Computer} {Science}},
	title = {On {Active} {Attack} {Detection} in {Messaging} with {Immediate} {Decryption}},
	volume = {14084},
	url = {https://doi.org/10.1007/978-3-031-38551-3\_12},
	doi = {10.1007/978-3-031-38551-3_12},
	booktitle = {Advances in {Cryptology} - {CRYPTO} 2023 - 43rd {Annual} {International} {Cryptology} {Conference}, {CRYPTO} 2023, {Santa} {Barbara}, {CA}, {USA}, {August} 20-24, 2023, {Proceedings}, {Part} {IV}},
	publisher = {Springer},
	author = {Barooti, Khashayar and Collins, Daniel and Colombo, Simone and Huguenin-Dumittan, Lo{\"i}s and Vaudenay, Serge},
	editor = {Handschuh, Helena and Lysyanskaya, Anna},
	year = {2023},
	pages = {362--395},
}

@inproceedings{eugster_cryptographic_2018,
	title = {A {Cryptographic} {Look} at {Multi}-party {Channels}},
	url = {https://doi.org/10.1109/CSF.2018.00010},
	doi = {10.1109/CSF.2018.00010},
	booktitle = {31st {IEEE} {Computer} {Security} {Foundations} {Symposium}, {CSF} 2018, {Oxford}, {United} {Kingdom}, {July} 9-12, 2018},
	publisher = {IEEE Computer Society},
	author = {Eugster, Patrick and Marson, Giorgia Azzurra and Poettering, Bertram},
	year = {2018},
	pages = {31--45},
}

@misc{barnes_messaging_2023,
	title = {The {Messaging} {Layer} {Security} ({MLS}) {Protocol}},
	url = {https://www.rfc-editor.org/info/rfc9420},
	doi = {10.17487/RFC9420},
	publisher = {RFC Editor},
	author = {Barnes, Richard and Beurdouche, Benjamin and Robert, Raphael and Millican, Jon and Omara, Emad and Cohn-Gordon, Katriel},
	month = jul,
	year = {2023},
	note = {Issue: 9420
Num Pages: 132
Series: Request for Comments
Published: RFC 9420},
}

@misc{beurdouche_messaging_2025,
	title = {The {Messaging} {Layer} {Security} ({MLS}) {Architecture}},
	url = {https://www.rfc-editor.org/info/rfc9750},
	doi = {10.17487/RFC9750},
	publisher = {RFC Editor},
	author = {Beurdouche, Benjamin and Rescorla, Eric and Omara, Emad and Inguva, Srinivas and Duric, Alan},
	month = apr,
	year = {2025},
	note = {Issue: 9750
Num Pages: 41
Series: Request for Comments
Published: RFC 9750},
}

@techreport{robert_messaging_2025,
	type = {Internet-{Draft}},
	title = {The {Messaging} {Layer} {Security} ({MLS}) {Extensions}},
	url = {https://datatracker.ietf.org/doc/draft-ietf-mls-extensions/08/},
	number = {draft-ietf-mls-extensions-08},
	institution = {Internet Engineering Task Force},
	author = {Robert, Raphael},
	month = jul,
	year = {2025},
	note = {Backup Publisher: Internet Engineering Task Force
Num Pages: 37},
}

@inproceedings{goldberg_multi-party_2009,
	title = {Multi-party off-the-record messaging},
	url = {https://doi.org/10.1145/1653662.1653705},
	doi = {10.1145/1653662.1653705},
	booktitle = {Proceedings of the 2009 {ACM} {Conference} on {Computer} and {Communications} {Security}, {CCS} 2009, {Chicago}, {Illinois}, {USA}, {November} 9-13, 2009},
	publisher = {ACM},
	author = {Goldberg, Ian and Ustaoglu, Berkant and Gundy, Matthew Van and Chen, Hao},
	editor = {Al-Shaer, Ehab and Jha, Somesh and Keromytis, Angelos D.},
	year = {2009},
	pages = {358--368},
}

@book{cachin_introduction_2011,
	title = {Introduction to {Reliable} and {Secure} {Distributed} {Programming} (2. ed.)},
	isbn = {978-3-642-15259-7},
	url = {https://doi.org/10.1007/978-3-642-15260-3},
	doi = {10.1007/978-3-642-15260-3},
	publisher = {Springer},
	author = {Cachin, Christian and Guerraoui, Rachid and Rodrigues, Lu{\'i}s E. T.},
	year = {2011},
}

@misc{robert_messaging_2025-1,
	title = {Messaging {Layer} {Security} ({MLS}): {Towards} {More} {End}-to-{End} {Encryption}},
	author = {Robert, Raphael},
	month = jul,
	year = {2025},
	note = {Published: Presentation, Pass the SALT 2025},
}

@inproceedings{schliep_end--end_2019,
	title = {End-to-{End} {Secure} {Mobile} {Group} {Messaging} with {Conversation} {Integrity} and {Deniability}},
	url = {https://www-users.cse.umn.edu/~hoppernj/cowpi_wpes19.pdf},
	doi = {10.1145/3338498.3358644},
	booktitle = {Proceedings of the 18th {ACM} {Workshop} on {Privacy} in the {Electronic} {Society}, {WPES}@{CCS} 2019, {London}, {UK}, {November} 11, 2019},
	publisher = {ACM},
	author = {Schliep, Michael and Hopper, Nicholas},
	editor = {Cavallaro, Lorenzo and Kinder, Johannes and Domingo-Ferrer, Josep},
	year = {2019},
	pages = {55--73},
}

@misc{lee_despite_2025,
	title = {Despite misleading marketing, {Israeli} company {TeleMessage}, used by {Trump} officials, can access plaintext chat logs},
	url = {https://micahflee.com/despite-misleading-marketing-israeli-company-telemessage-used-by-trump-officials-can-access-plaintext-chat-logs/},
	author = {Lee, Micah},
	month = may,
	year = {2025},
}

@misc{goldberg_trump_2025,
	title = {The {Trump} {Administration} {Accidentally} {Texted} {Me} {Its} {War} {Plans}},
	url = {https://www.theatlantic.com/politics/archive/2025/03/trump-administration-accidentally-texted-me-its-war-plans/682151},
	publisher = {The Atlantic},
	author = {Goldberg, Jeffrey},
	month = mar,
	year = {2025},
}

@misc{holland_eu_2025,
	title = {{EU} {Signal} {Group}: {Too} sensitive for release, not sensitive enough for archiving},
	url = {https://www.heise.de/en/news/EU-Signal-Group-Too-sensitive-for-release-not-sensitive-enough-for-archiving-11080698.html},
	publisher = {heise online},
	author = {Holland, Martin},
	month = nov,
	year = {2025},
}

@article{schliep_consistent_2018,
	title = {Consistent {Synchronous} {Group} {Off}-{The}-{Record} {Messaging} with {SYM}-{GOTR}},
	volume = {2018},
	url = {https://doi.org/10.1515/popets-2018-0027},
	doi = {10.1515/POPETS-2018-0027},
	number = {3},
	journal = {Proc. Priv. Enhancing Technol.},
	author = {Schliep, Michael and Vasserman, Eugene Y. and Hopper, Nicholas},
	year = {2018},
	pages = {181--202},
}

@inproceedings{bracha_asynchronous_1984,
	address = {Vancouver, British Columbia, Canada},
	title = {An {Asynchronous} [(n-1)/3]-{Resilient} {Consensus} {Protocol}},
	url = {https://dl.acm.org/doi/10.1145/800222.806743},
	doi = {10.1145/800222.806743},
	booktitle = {Proceedings of the {Third} {Annual} {ACM} {Symposium} on {Principles} of {Distributed} {Computing} ({PODC} '84)},
	publisher = {Association for Computing Machinery},
	author = {Bracha, Gabriel},
	year = {1984},
	pages = {154--162},
}

@inproceedings{toueg_randomized_1984,
	address = {Vancouver, British Columbia, Canada},
	title = {Randomized {Byzantine} {Agreements}},
	url = {https://dl.acm.org/doi/10.1145/800222.806744},
	doi = {10.1145/800222.806744},
	booktitle = {Proceedings of the {Third} {Annual} {ACM} {Symposium} on {Principles} of {Distributed} {Computing} ({PODC} '84)},
	publisher = {Association for Computing Machinery},
	author = {Toueg, Sam},
	year = {1984},
	pages = {163--178},
}

@inproceedings{martiny_improving_2021,
	title = {Improving {Signal}'s {Sealed} {Sender}},
	url = {https://www.cs.umd.edu/~kaptchuk/publications/ndss21.pdf},
	booktitle = {28th {Annual} {Network} and {Distributed} {System} {Security} {Symposium}, {NDSS} 2021, virtually, {February} 21-25, 2021},
	publisher = {The Internet Society},
	author = {Martiny, Ian and Kaptchuk, Gabriel and Aviv, Adam J. and Roche, Daniel S. and Wustrow, Eric},
	year = {2021},
}

@inproceedings{pass_sleepy_2017,
	series = {Lecture {Notes} in {Computer} {Science}},
	title = {The {Sleepy} {Model} of {Consensus}},
	volume = {10624},
	doi = {10.1007/978-3-319-70697-9_14},
	booktitle = {Advances in {Cryptology} {\textendash} {ASIACRYPT} 2017},
	publisher = {Springer},
	author = {Pass, Rafael and Shi, Elaine},
	year = {2017},
	pages = {380--409},
}

@article{dwork_consensus_1988,
	title = {Consensus in the {Presence} of {Partial} {Synchrony}},
	volume = {35},
	doi = {10.1145/42282.42283},
	number = {2},
	journal = {Journal of the ACM},
	author = {Dwork, Cynthia and Lynch, Nancy and Stockmeyer, Larry},
	year = {1988},
	pages = {288--323},
}

\appendix
\section{Appendix}
\subsection{Defining Transcript Consistency}
\label{sec:TC}

The ambiguous use of the term \emph{transcript consistency}, e.g. in the context of MLS where it is only referring to group state~\cite{barnes_messaging_2023,beurdouche_messaging_2025,robert_messaging_2025-1}, as well as the various informal descriptions of its desired properties~\cite{goldberg_multi-party_2009,kleppmann_secure_2018,marlinspike_private_2014} necessitate more fine-grained definitions of the different flavors of this security property.

We define four versions of this security property in the context of secure group messaging, allowing for different design and implementation approaches.
Our definitions range from the weakest form of TC to its strongest form. 
For all versions, we consider a group chat with a set of participants $\mathcal{P}={ P_1, P_2, P_3, \dots, P_n }$ that are already added to the group\footnote{Users may operate multiple linked devices, each maintaining independent cryptographic state, as supported by the messaging systems under investigation. At a high level, this behavior can be modeled as separate participants.}.
The attacker is a member of the respective group conversation and controls at most $ f $ participants.
The total group size is fixed and denoted by $ n $, where $ f \le \left\lfloor \frac{n - 1}{3} \right\rfloor$ holds.
This bound is necessary to ensure that agreement on a common message transcript is theoretically possible under asynchrony and malicious  participants~\cite{bracha_asynchronous_1984,toueg_randomized_1984}.

Communication in the context of instant messaging is assumed to be asynchronous. Moreover, honest participants can go offline at any time, akin to the sleepy model in consensus~\cite{pass_sleepy_2017}. 
For TC this is challenging, as a consistent transcript cannot be ensured for all honest participants in times where honest participants are offline. We account for this in our definitions of TC by omitting when exactly, or how often, TC should be achieved. This leaves room for different approaches as to when TC should be verified or enforced.
In a basic form, it could be sufficient to \emph{eventually} arrive at  a consistent transcript: When all participants are online and no additional message is sent within the group for at least $\Delta$ time, where $\Delta$ is the propagation delay (network as well as processing) of the slowest participant in a group\footnote{This assumption is comparable to GST (Global Stabilization Time)~\cite{dwork_consensus_1988}.}. 
Therefore, TC can eventually be achieved if such situations arise. 
This deliberately leaves unspecified whether participants are required to know when TC is achieved, or whether it is sufficient to know that TC remains achievable and has not been violated.

\begin{definition}[Set Transcript Consistency]
\label{def:set-transcript-consistency}
Consider a group of participants $\mathcal{P}$ executing a secure
group messaging protocol in the presence of an adversary that controls a subset of Byzantine participants.
Each participant $i \in \mathcal{P}$ maintains a local \emph{transcript}
$\mathsf{Tr}_i$, defined as the collection of messages that $i$ accepts
as belonging to the conversation after performing all protocol-defined
verification checks.

The protocol provides \emph{Set Transcript Consistency (STC)} if, for any
execution and for any two honest participants $i,j \in P$, the following
eventually holds:
\[
\mathrm{set}(\mathsf{Tr}_i) = \mathrm{set}(\mathsf{Tr}_j).
\]

That is, all honest participants eventually accept exactly the same set of messages, independently of the order in which those messages are delivered or displayed locally.
\end{definition}

STC should reflect the first description of such a property by Goldberg et al.~\cite{goldberg_multi-party_2009} which suggested to perform this consistency check (termed \emph{consensus} in~\cite{goldberg_multi-party_2009}) in retrospect after shutting down a group chat by lexicographically ordering all messages and comparing the digests.

\begin{definition}[Participant Transcript Consistency]
\label{def:speaker-transcript-consistency}
Consider a group of participants $\mathcal{P}$ executing a secure group messaging protocol in the presence of an adversary that controls a subset of Byzantine participants.
Each participant $i \in \mathcal{P}$ maintains a local \emph{transcript}
$\mathsf{Tr}_i$, defined as the set of messages that $i$ accepts as belonging to the conversation after performing all protocol-defined verification checks.

Each message $m$ is associated with a unique sender
$\mathsf{snd}(m) \in \mathcal{P}$ and a sender-local sequence number $\mathsf{seq}(m) \in \mathbb{N}$.

The protocol provides \emph{Participant Transcript Consistency (PTC)} if, for any execution and for any two honest participants $i,j \in \mathcal{P}$, the
following properties eventually hold:

\begin{enumerate}
  \item \textbf{Set Agreement:}
  \[
  \mathrm{set}(\mathsf{Tr}_i) = \mathrm{set}(\mathsf{Tr}_j).
  \]

  \item \textbf{Per-Speaker FIFO Order:}
  For any two messages $m_1,m_2$ such that
  $\mathsf{snd}(m_1) = \mathsf{snd}(m_2)$ and
  $\mathsf{seq}(m_1) < \mathsf{seq}(m_2)$, if $m_1$ precedes $m_2$ in
  $\mathsf{Tr}_i$ for some honest participant $i$, then $m_1$ precedes
  $m_2$ in $\mathsf{Tr}_j$ for every honest participant $j$.
\end{enumerate}

No ordering constraint is imposed on messages originating from different
senders.
\end{definition}

This definition of PTC is comparable to the property termed \emph{speaker consistency} by Unger et al.~\cite{unger_sok_2015}.

\begin{definition}[Causal Transcript Consistency]
\label{def:causal-transcript-consistency}
Consider a group of participants $\mathcal{P}$ executing a secure group messaging protocol in the presence of an adversary that controls a subset of Byzantine participants.
Each participant $i \in \mathcal{P}$ maintains a local \emph{transcript}
$\mathsf{Tr}_i$, defined as the set of messages that $i$ accepts as belonging to the conversation after performing all protocol-defined verification checks, together with a local delivery order.

Let $\rightarrow$ denote a protocol-defined \emph{causal precedence} relation over messages (e.g., induced by send-receive dependencies or explicit causal metadata).

The protocol provides \emph{Causal Transcript Consistency (CTC)} if, for any execution and for any two honest participants $i,j \in \mathcal{P}$, the following properties eventually hold:

\begin{enumerate}
  \item \textbf{Set Agreement:}
  \[
  \mathrm{set}(\mathsf{Tr}_i) = \mathrm{set}(\mathsf{Tr}_j).
  \]

  \item \textbf{Causal Order Agreement:}
  For any two messages $m_1,m_2$ such that $m_1 \rightarrow m_2$, if
  $m_1 \rightarrow m_2$ in $\mathsf{Tr}_i$ for some honest participant $i$,
  then $m_1 \rightarrow m_2$ in $\mathsf{Tr}_j$ for every honest
  participant $j$.
\end{enumerate}
\end{definition}

This definition of TC captures a form of causal ordering of messages: If a message $m_1$ was received before a message $m_2$ was sent, then $m_2$ causally depends on $m_1$, i.e., $m_1 \rightarrow m_2$, and must be delivered after $m_1$. Causal ordering only relates any two messages, but does not allow putting every message in relation to any other message, s.t. messages can also be at the same position. %

\aju{The causal ordering allows for elements to be independent/at the same time. The receive and sent property depends on who is the broadcaster and how the messages are collected. The orderings of different participants might be different.

Rough concept: Server BCBs messages and the clients receive the messages and deliver them in a causal order. This does not put any constraints on how the server is collecting these messages.}

\begin{definition}[Total Transcript Consistency]
\label{def:total-transcript-consistency}
Consider a group of participants $\mathcal{P}$ executing a secure group messaging protocol in the presence of an adversary that controls a subset $ f $ of Byzantine participants.
Each participant $i \in \mathcal{P}$ maintains a local \emph{transcript}
$\mathsf{Tr}_i$, defined as the sequence of messages that $i$ accepts
as belonging to the conversation after performing all protocol-defined verification checks.

The protocol provides \emph{Total Transcript Consistency (TTC)} if, for any execution and for any two honest participants $i,j \in \mathcal{P}$, the following eventually holds:
\[
\mathsf{Tr}_i = \mathsf{Tr}_j .
\]

Equivalently, for any two messages $m_1$ and $m_2$ that appear in the transcripts of honest participants, if $m_1$ precedes $m_2$ in $\mathsf{Tr}_i$ for some honest participant $i$, then $m_1$ precedes $m_2$ in $\mathsf{Tr}_j$ for every honest participant $j$.
\end{definition}

TTC should capture the informal notion of \emph{global transcript}~\cite{unger_sok_2015,kleppmann_secure_2018}, \emph{transcript agreement}~\cite{cohn-gordon_ends--ends_2018} as well as the intuitive meaning described by Marlinspike~\cite{marlinspike_private_2014}.

\dsc{It is the current section. Thus I comment the next paragraph out}

\aju{Public note to self: The tricky question is when TC should be checked? In retrospect or continuously assured? This kind of also related to a property termed instant delivery in the context of secure instant messengers}

\subsection{Signal: Inconsistent Duplicate Handling}
\label{sec:appendix:signal-inconsistent-duplicate-handling}
During our experiments, we noticed additional implementation differences between Signal's
mobile app and the desktop client. Duplicates in Signal are usually detected if a message
with the same author service ID and client timestamp has been received before. The desktop
client, however, also takes the sender's device ID into account, making it possible to receive
two messages from different devices with the same sender service ID and client timestamp pair.
This is demonstrated in the artifact video~\cite{signal_video_15}. %

Another possibility to receive a duplicate message with Signal's desktop client is if the edit
of a message has the same timestamp as a message sent after this edit. The message with the
reused timestamp is dropped by the official Android client, for example, but accepted by
the desktop client, as shown by the artifact video~\cite{signal_video_16}.%

\end{document}